\documentclass{aa}  

\usepackage{graphicx}
\usepackage{txfonts}
\usepackage[]{hyperref}
\begin{document}

   \title{An all-sky sample of intermediate- to high-mass OBA-type eclipsing binaries observed by TESS\thanks{The two catalogue files of OBA stars and of eclipsing binaries are available at the CDS via anonymous ftp to \url{cdsarc.u-strasbg.fr} (130.79.128.5) or via \url{http://cdsweb.u-strasbg.fr/}}}

   \subtitle{}

   \author{L. W. IJspeert\inst{1}
          \and
          A. Tkachenko\inst{1}
          \and
          C. Johnston\inst{1, 2}
          \and
          S. Garcia\inst{1}
          \and
          J. De Ridder\inst{1}
          \and
          T. Van Reeth\inst{1}
          \and
          C. Aerts\inst{1, 2, 3}
          }

   \institute{Institute of Astronomy, KU Leuven,
              Celestijnenlaan 200D, 3001 Leuven, Belgium\\
              \email{luc.ijspeert@kuleuven.be}
        \and
              Department of Astrophysics, IMAPP, Radboud University Nijmegen, 
              P. O. Box 9010, 6500 GL Nijmegen, the Netherlands\\
        \and
              Max Planck Institute for Astronomy, 
              K\"onigstuhl 17, 69117 Heidelberg, Germany
             }

   \date{Received June 08, 2021; accepted July 13, 2021}

 
  \abstract
   {Intermediate- to high-mass stars are the least numerous types of stars and they are less well understood than their more numerous low-mass counterparts in terms of their internal physical processes. Modelling the photometric variability of a large sample of main-sequence intermediate- to high-mass stars in eclipsing binary systems will help to improve the models for such stars.}
   {Our goal is to compose a homogeneously compiled sample of main-sequence intermediate- to high-mass OBA-type dwarfs in eclipsing binary systems from TESS photometry. We search for binaries with and without pulsations and determine their approximate ephemerides.}
   {Our selection starts from a catalogue of dwarfs with colours corresponding to those of OBA-type dwarfs in the TESS Input Catalog. We develop a new automated method aimed at detecting eclipsing binaries in the presence of strong pulsational and/or rotational signal relative to the eclipse depths and apply it to publicly available 30-min cadence TESS light curves.}
   {Using targets with TESS magnitudes below 15 and cuts in the 2MASS magnitude bands of $J - H < 0.045$ and $J - K < 0.06$ as most stringent criteria, we arrive at a total of 189\,981 intermediate- to high-mass candidates, 91193 of which have light curves from at least one of two data reduction pipelines. The eclipsing binary detection and subsequent manual check for false positives resulted in 3155 unique OBA-type eclipsing binary candidates.}
   {Our sample of eclipsing binary stars in the intermediate- to high-mass regime allows for future binary (and asteroseismic) modelling with the aim to better understand the internal physical processes in this hot part of the main sequence.}


   \maketitle
%

\section{Introduction}

The knowledge of stellar interiors has come a long way for many types of stars across the Hertzsprung Russell diagram, leading to an overall proper theory of stellar structure and evolution \citep{Kippenhahn2012}. 
In-depth asteroseismic analyses of large samples of low-mass dwarfs have shown their internal structure to be fairly similar to the one of the Sun \citep[][for a review]{GarciaBallot2019}. The largest samples of low-mass stars with asteroseismology have been gathered for red giants, whose mixed dipole modes allowed to pinpoint their evolutionary stage \citep{Bedding2011} yet revealed shortcomings in the theory of angular momentum transport \citep{Beck2012, Mosser2014, Gehan2018, Aerts2019}. 

At the high mass end, asteroseismic studies are scarce due to low numbers of such stars observed with high-precision uninterrupted space photometry and due to the challenge of mode identification \citep{Aerts2010, 2013pss4.book..207H, Bowman2020c}. Yet, stellar evolution models are most uncertain for the most massive stars and, moreover, these stars are dominant in the production of heavy elements \citep{Hirschi2005}. 
So far, only a modest sample of intermediate-mass B-type stars has been scrutinised for internal rotation and mixing properties by asteroseismology \citep{Pedersen2021}. This study revealed the envelope mixing properties of rotating B-type stars to be very different from those of low-mass stars with slow rotation \citep[][Table\,1]{Aerts2021}. 
Asteroseismic probing of the internal rotation and mixing physics in samples of O- and B-type stars with masses in the regime of supernova progenitors remain scarce. Few individual $\beta\,$Cep stars have been modelled from ground- or space-based asteroseismology and those studies were done in an inhomogeneous way \citep[e.g.,][]{Dupret2004, Briquet2007, Dziembowski2008, Briquet2012, Daszynska2017, Handler2019}.

The two major barriers towards a general asteroseismic understanding of the interior processes in the high-mass O- and B-type stars are small sample size and lack of proper mode identification. 
The previous large-scale CoRoT \citep{2009A&A...506..411A} and \textit{Kepler} \citep{2010ApJ...713L..79K} surveys resulted in a relatively small set of OB-type stars with high precision photometric time series measurements from space and almost none of the O-type pulsators came with unambiguous mode identification \citep[see Table\,3 in][]{2015MNRAS.453...89B}. Notable exceptions delivering asteroseismic inferences of an O9V and two early Be stars are described in \citet{Briquet2011}, \citet{Neiner2012}, and \citet{Neiner2020}, respectively. The latter two Be stars were found to have gravito-inertial modes excited by the convective core \citep{Neiner2020}, following the theory of gravito-inertial wave generation by \citet{Augustson2020}. Internal gravity waves excited by core convection have also been suggested as a physical explanation for the detected spectra of low-frequency stochastic variability found in large samples of intermediate- and high-mass stars observed with CoRoT, \textit{Kepler}\hyphenation{Kep-ler}, \textit{K2}, and TESS \citep{Bowman2019, 2019NatAs...3..760B, 2020A&A...640A..36B, 2021MNRAS.502.5038N, Szewczuk2021}. While such stochastically excited wave spectrum has not yet been used to probe internal physics, it is compatible with the predictions from large-scale multi-dimensional numerical simulations covering hundreds of convective turnover time scales \citep{Edelmann2019,Horst2020,Rathish2020}. Bridging such simulations and observations may open up asteroseismology based on stochastic low-frequency gravito-inertial waves, provided that the degree and azimuthal order of resonant modes occurring within the excited broad wave spectrum can be identified. 

Clearly, early-type stars demand dedicated effort to improve their interior physics from asteroseismology because those with identified modes are few in number among their lower-mass counterparts. This is mainly due to the nature of the Initial Mass Function and because only a limited number of them provide us with proper oscillation modes that allow for seismic inferences.
The NASA Transiting Exoplanet Survey Satellite (TESS) mission \citep{2015JATIS...1a4003R} is supplementing the existing time series space photometry with a treasure trove of data covering a large portion of the sky, including many intermediate- to high-mass stars. Initial searches for optimal OB-type pulsators have been conducted \citep{Pedersen2019, Burssens2020}. 
Here, we take an entirely complementary and systematic approach, by exploiting the unique TESS all-sky data set in a hunt for intermediate- to high-mass OBA-type eclipsing binaries with the aim to compose an overall optimal sample of stars for stellar modelling.

To study stellar interiors through forward binary and/or asteroseismic modelling, it is highly beneficial to constrain as many observables as possible, including model-independent parameters, given the high dimensionality of the parameter estimation problem at hand \citep{Prsa2016,Aerts2021}.
Dynamical masses and mass ratios deduced for binary systems offer a good diagnostic to treat challenging discrepancies between models and observations \citep[e.g.][]{Tkachenko2020}. Moreover, the complementarity between binarity and stellar oscillations offers an optimal way forward to binary asteroseismology  \citep{2019MNRAS.482.1231J}. For this purpose, the approach presented here focuses on eclipsing binary (EB) stars as these present the opportunity to measure model-independent stellar masses and radii. 
While searches for individual pulsating OB-type EB targets \citep{Southworth2020, Southworth2021}, as well as medium-size samples of interesting targets like massive heartbeat stars \citep{2021A&A...647A..12K}, are ongoing, we take a homogeneous systematic large-scale approach. A notable large-scale sample of TESS eclipsing binaries in the Southern Hemisphere that covers a wide range in temperatures using a systematic selection procedure has been compiled by \citet{2021ApJ...912..123J}. Here, we limit our search to colours for OB dwarfs but we consider the full sky covered by TESS and aim to compose a sample with asteroseismic potential. This approach will lead to far less candidates but, fortunately, as a counterbalance to the Initial Mass Function, O- and B-type stars have a high binarity rate \citep{Moe2017} such that numerous detections of new EBs are anticipated.

This work presents a dedicated effort to find suitable EB candidates for stellar modelling in the intermediate- to high-mass OBA-type regime from TESS all-sky photometric data. We introduce a new target selection method specifically aimed at finding eclipses in light curves that might also have intrinsic sources of variability in them, most importantly stellar oscillations. We specifically aim to compose a large sample of pulsating and non-pulsating intermediate- to high-mass EBs. It has been shown that missing physics in terms of angular momentum transport already occurs during the main-sequence phase \citep{Aerts2019}. For this reason, and to delimit our science aims, we search detached (meaning neither components are filling their Roche lobe) main-sequence EBs with an intermediate- to high-mass primary. This is in part also for practical reasons, since evolved stars in this mass regime have strong stellar winds that complicate the interpretation of their light curves and the modelling of their interiors.

We seek to find EB candidates with clear secondary eclipses and large enough time base for precise modelling as we want to compose a sample that is optimally suited to calibrate stellar evolution models. Characterisation of ensembles of stars that show stellar oscillations versus those that do not will give insights in the role and the importance of these oscillations in the transport of angular momentum and chemical elements. Comparison between ensembles of binaries experiencing strong tidal forces versus binaries that evolve as if they were single stars will provide a handle on the role and importance of tidal forces in the internal processes. These considerations have guided our search described in the rest of the paper.

\section{Selection of intermediate- to high-mass candidates}
\label{selection}

We start off by producing an all-sky catalogue of intermediate- to high-mass candidates from the TESS Input Catalog (TIC) \citep{2019AJ....158..138S}, which serves as the initial selection for our intermediate- to high-mass EB candidates. We use the latest version of the TIC (TICv8), which relies on Gaia DR2 \citep{2018A&A...616A...1G} as its base catalogue. From there, we proceed by analysing the publicly released 30-minute cadence data from the first and second year of TESS observations (sectors 1-26) to arrive at the presented sample of EBs. This includes the northern and southern continuous viewing zones (CVZs), providing the most promising candidates for modelling. Indeed, the duration of the photometric time series is an important factor in the asteroseismic exploitation of the observations, since the frequency resolution scales with the inverse of the time base for coherent modes \citep[][Chapter\,5]{Aerts2010}. Additionally, covering more than few eclipses improves the precision on the fit of the EB model, particularly in the presence of other sources of variability. 

Overall, we present a catalogue of intermediate- to high-mass EB candidates based on the presence of eclipses in the TESS photometry, including their approximate ephemerides. The byproduct of this is a catalogue of intermediate- to high-mass candidates that we also make publicly available through CDS.

\begin{figure}
\centering
\includegraphics[width=\hsize]{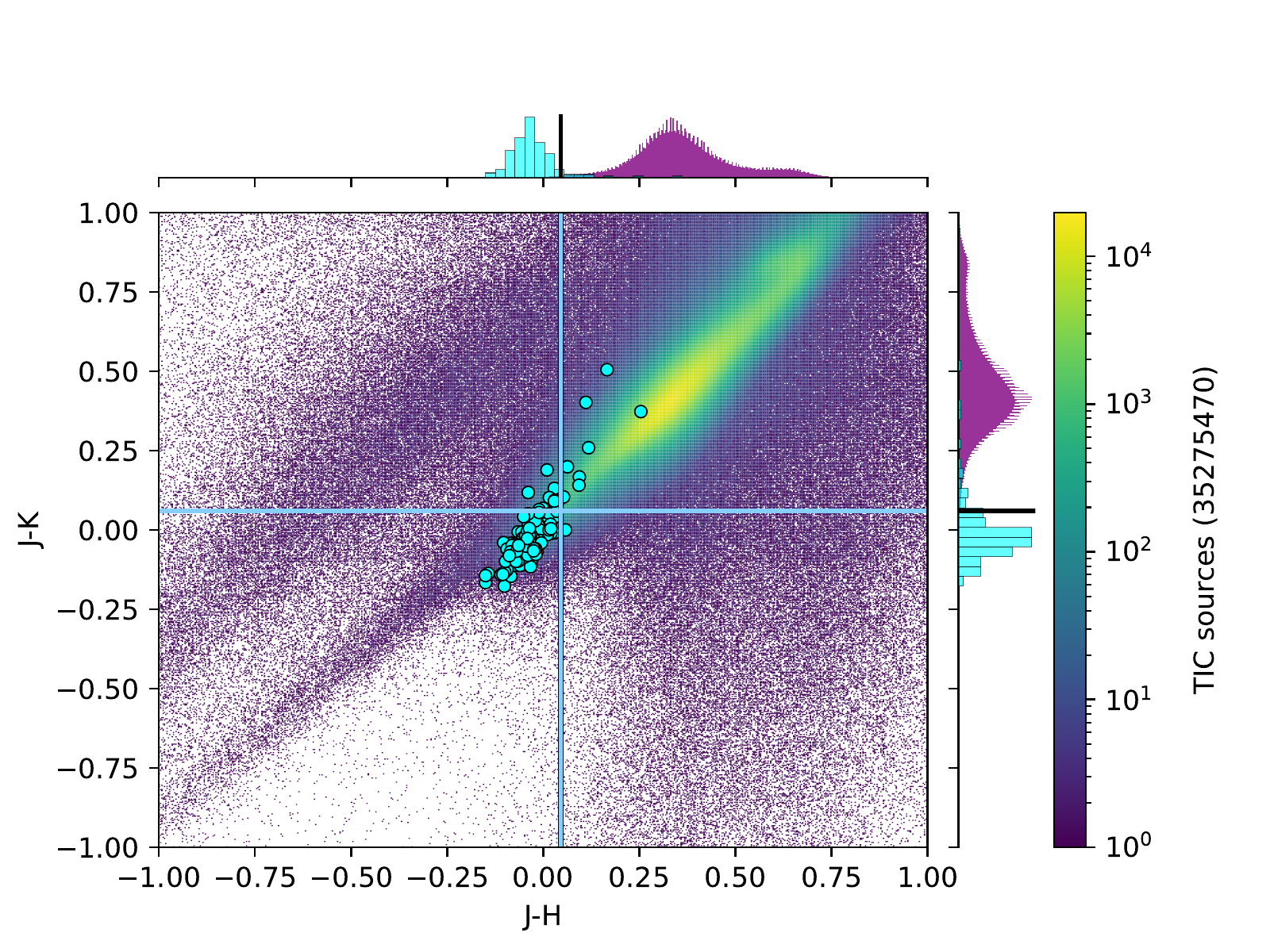}
  \caption{Colour-colour plot of the \citet{Pedersen2019} (cyan circles) sample over a background of TIC targets brighter than 15th magnitude. The colour cuts are indicated by the horizontal and vertical lines. Only targets in the lower left quadrant are kept.}
     \label{fig:colour}
\end{figure}

\subsection{Selection criteria}
\label{sec:ob_sel}

We limit ourselves to TESS magnitudes below 15 and targets that have the flag \texttt{objType == `STAR'} to filter out extended objects, which reduces the number of targets from the total size of the TIC of $1.7\cdot 10^9$ targets to $6.6\cdot 10^7$. We further exclude the objects for which only one target occurs in the 2MASS catalogue but two targets are listed in the Gaia catalogue, using the flag \texttt{disposition != `SPLIT'}. Such targets are inherently contaminated by the light of the second source, which cannot be distinguished, and thus lead to wrong conclusions if these targets are analysed as if only one source is responsible for the photometric time series. The cut in the number of sources is small and brings us to $6.55\cdot 10^7$ targets (-0.82\%). Subsequently we retain only those targets having a measurement in each of the three 2MASS JHK near-infrared photometric bands \citep{2006AJ....131.1163S}. We use those 2MASS measurements to construct the colours $J-H$ and $J-K$ for each target. This further reduces the number of targets to $6.37\cdot 10^7$ (-2.7\%).

To select the appropriate temperature regime, a suitable colour cut is made using the 2MASS colours and adopting the results from the manual target selection used by \citet{Pedersen2019}. They selected a sample of 154 O- and B-type stars using the spectral types provided on \texttt{SIMBAD}\footnote{\url{http://simbad.u-strasbg.fr/simbad/}}.
The colour cut is determined by placing this sample in a colour-magnitude diagram (CMD) and excluding too red targets (likely due to reddening), in order to minimise the contamination by cooler stars (see Fig.\,\ref{fig:colour}). Two cuts are made: $J - H < 0.045$ and $J - K < 0.06$. This brings the sample down to 205\,936 targets (a reduction of 99.7\%). Fig.\,\ref{fig:colour} shows the sample by \citet{Pedersen2019} in cyan plotted over a density map of TIC targets around the intersection of the colour cuts indicated by the vertical and horizontal lines.

The use of infrared filters minimises the effect of reddening; these colour cuts are conservative in the sense that correcting for reddening would increase the number of retained targets. Using these colour cuts less affected by reddening is a trade-off with temperature sensitivity, as the tail of the spectral energy distribution for the stars of interest flattens off towards the red. Ideally we would include an absolute magnitude cut as well, but the parallaxes from the Gaia space satellite \citep{2016A&A...595A...1G, 2018A&A...616A...1G} required for this are not yet available for binary systems. For the blue section of the CMD that we are looking at, the gain of adding such an additional cut is small in terms of contamination.

Since we are interested in main-sequence stars, we subsequently applied the additional flags \texttt{wdflag != 1} and \texttt{lumclass != `GIANT'} to all 205\,936 targets in order to filter out white dwarfs and giants. This reduces the sample by a further 7.7\%. The final number of targets in the intermediate- to high-mass OBA-type candidate sample is 189\,981.

\begin{figure}
\centering
\includegraphics[width=\hsize]{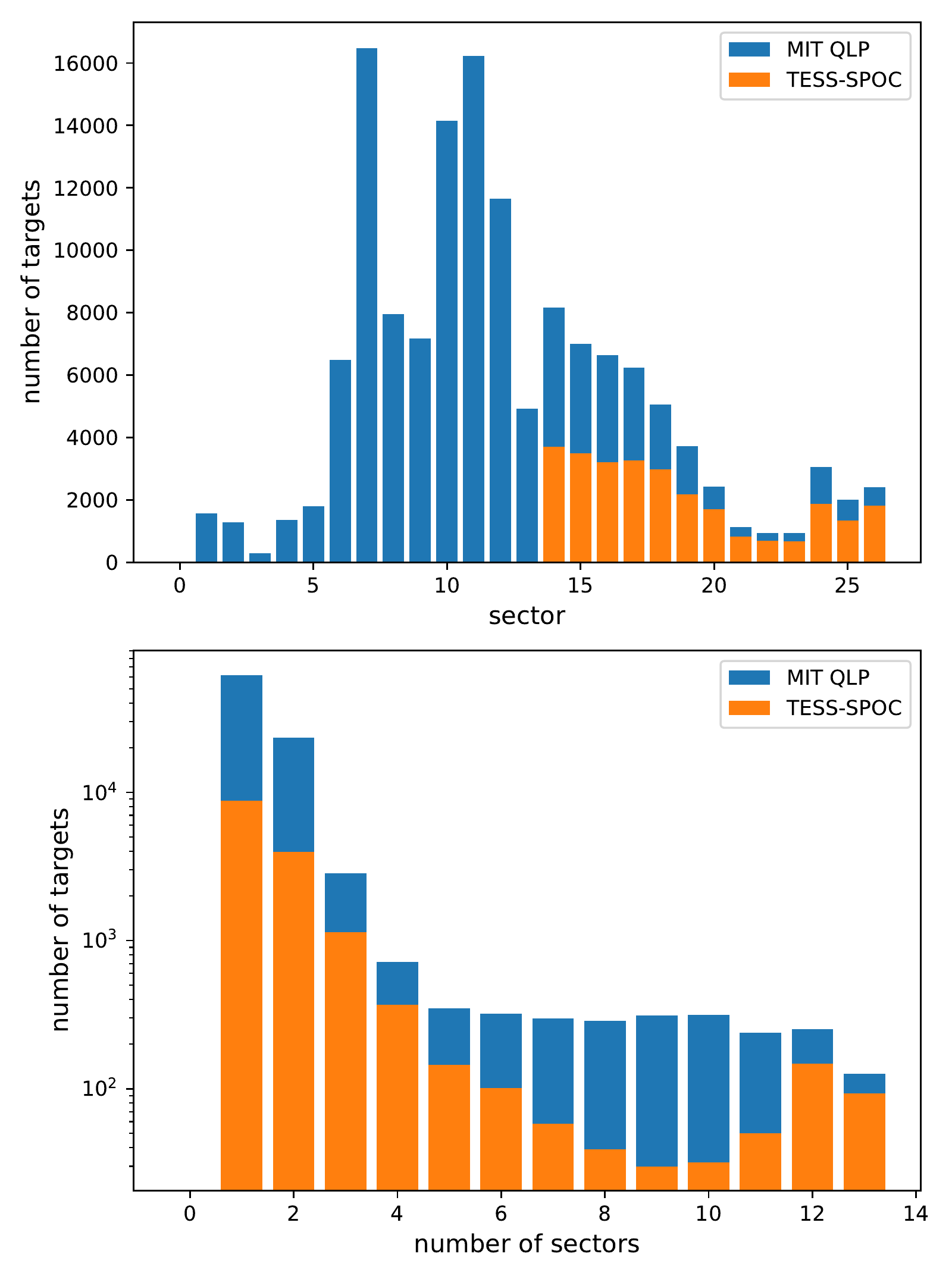}
  \caption{(top) Distribution of targets over the TESS sectors of the first and second year. (bottom) The distribution of the number of sectors per target. Most targets have one or two sectors of observations.}
     \label{fig:sectors}
\end{figure}

\subsection{TESS light curves}
\label{sec:eb_sel}

For the next step we used light curves from the public data releases of 30-min cadence data processed by TESS-SPOC (the TESS Science Processing Operations Center)\footnote{\href{https://dx.doi.org/10.17909/t9-wpz1-8s54}{DOI: 10.17909/t9-wpz1-8s54}} and by the MIT QLP (the MIT Quick-Look Pipeline)\footnote{\href{https://dx.doi.org/10.17909/t9-r086-e880}{DOI: 10.17909/t9-r086-e880}}. We focused on observations from the nominal TESS mission available in the Mikulski Archive for Space Telescopes (MAST)\footnote{\url{http://archive.stsci.edu/tess/all\_products.html}} downloaded on 10 March 2021. This means sectors 1-13 (the ecliptic south) only have coverage by MIT QLP data and sectors 14-26 (the ecliptic north) have coverage by two data sets resulting from both pipelines. 
Cross-matching our OB candidate sample with the available light curves results in a total of 14970 available light curves from the TESS-SPOC and 91142 from the MIT QLP. The total number of unique targets with a light curve is 91193, so there is an almost complete overlap.

The TESS space telescope has a field of view of 24 by 96 degrees and it observes a portion of the sky for 27.3\,d before turning to a new  position. This is referred to as a sector. Over the time span of a year, 13 sectors cover a celestial hemisphere, leaving a relatively small area unobserved. There is an overlap of sectors at the ecliptic polar regions, offering a year of continuous coverage. Figure\,\ref{fig:sectors} shows how the light curves from our sample are distributed over the different sectors (top panel), and how many sectors are available for each target (bottom panel).

\section{Finding eclipsing binaries}

Several methods exist to detect eclipses in the light curves of stars, for instance the Box-fitting Least Squares (BLS) method \citep{2002A&A...391..369K}, Transit Least Squares (TLS) \citep{2019A&A...623A..39H}, or fitting eclipses locally to the light curve \citep{2019ApJS..244...11K}. Alternatively, one can use Fourier analysis of sinusoidal components, for example based on the use of Lomb-Scargle periodograms \citep{1976Ap&SS..39..447L, 1982ApJ...263..835S} suitable for non-uniformly sampled data. 

The BLS method works particularly well in low signal-to-noise cases where the eclipse depth is similar to the noise of the individual data points. This method is geared towards finding transit\footnote{The flat-bottomed eclipses of a smaller object occulting its host star.}-like features in an otherwise flat light curve. This means that in cases where the eclipse depths are similar to or smaller than the amplitude of other variability (e.g. pulsations or rotational modulation), it is nearly impossible to disentangle the eclipse signal in the time domain without first removing the other signal. Searching eclipses in the time domain in this way is also relatively time consuming. More advanced methods aimed specifically at exoplanet transits implementing Machine Learning methods include \citet{2018AJ....155...94S} and \citet{2021ApJ...912..123J}. In case of searches via Fourier analysis, eclipse signals and intrinsic variability frequencies are superimposed in the frequency spectrum. A high relative amplitude of for example pulsations with respect to eclipse depths may make the ability to detect eclipses difficult. 
\citet{2011A&A...529A..89D} use Gaussian mixtures for general variability classification, relying on the three most dominant frequencies in an LS periodogram, as well as their harmonics. The authors additionally implemented an extractor-type selection based on a high-pass filter to isolate EBs from all other types of variable stars. Such a filter removes the low-frequency pulsations typical for $\gamma$ Dor and SPB stars, after which the light curve is checked for downward outliers. \citet{2016MNRAS.456.2260A}, on the other hand, use self-organising maps to isolate the EBs.

Our aim is specifically to find pulsating EBs as well as non-pulsating ones. In order to achieve this in an efficient way for a large number of stars, we developed an algorithm to find eclipses in light curves that might also show other forms of variability. This algorithm, named \texttt{ECLIPSR} (\textit{Eclipse Candidates in Light curves and Inference of Period at a Speedy Rate}), has two main stages of operation: finding eclipse candidates in the light curve and determining the (orbital) period. It is made publicly available through \href{https://github.com/LucIJspeert/eclipsr}{GitHub}\footnote{\url{https://github.com/LucIJspeert/eclipsr}}. The following sections give an overview of how \texttt{ECLIPSR} works.

\begin{figure}
\centering
\includegraphics[width=\hsize]{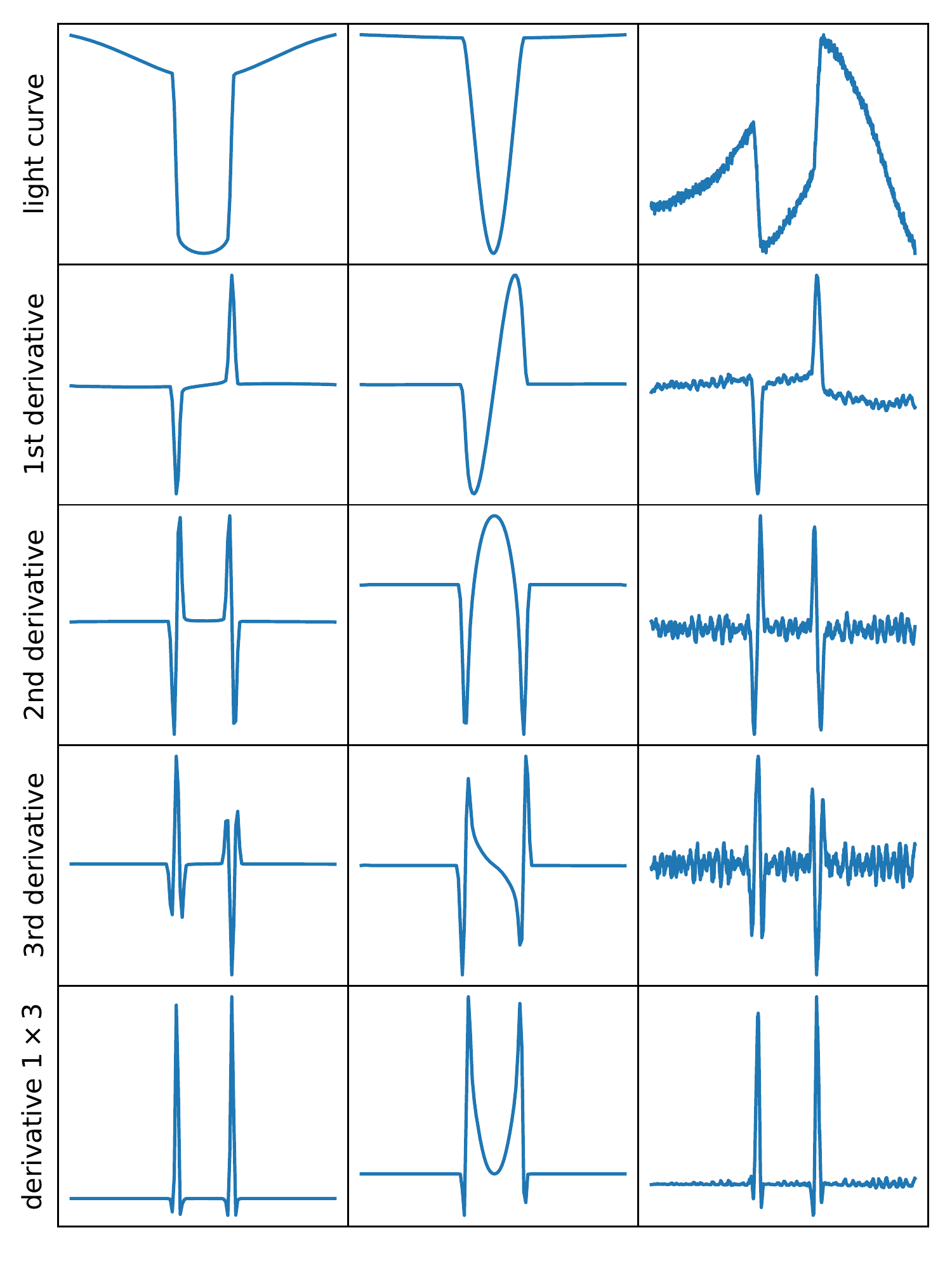}
  \caption{Typical eclipse signatures in the derivatives of the light curve. The left column shows a flat-bottomed eclipse resulting in two separated features in the derivatives, while the V-shaped eclipse in the middle leads to peaks merged into one feature. The right column shows an example with noise, which is an eclipse from TIC 93911780.}
     \label{fig:derivs}
\end{figure}

\subsection{Marking candidate eclipses}

We make a morphological distinction between two types of eclipses: flat-bottomed and V-shaped. Flat-bottomed eclipses (left and right columns of Fig.\,\ref{fig:derivs}) occur when either totality occurs for an extended period of time, or in the case of a transit of a smaller body in front of a larger one. The V-shaped eclipses (middle column of Fig.\,\ref{fig:derivs}) typically occur in the case of partial eclipses.
A third term that we use is `full eclipses'. This refers to eclipses with both ingress and egress detected in the data, which the algorithm determined belong together and thus combined to form one whole.

Steeper slopes occur at the eclipse ingress and egress compared to the other regions of the light curve for typical EB light curves. This shows up as positive and negative values for the first time derivative of the light curve. Eclipses of detached binaries show a distinctive strong drop or `bend' in the light curve at the start and end of the eclipse, implying steep first derivatives and high values for the second derivative. The third derivative resembles a horizontally mirrored version of the first derivative, with additional bumps to either side and peaks are narrower overall. Figure\,\ref{fig:derivs} shows a schematic overview of the characteristic eclipse signal in each derivative for a flat-bottomed and a V-shaped eclipse.

\begin{figure*}
\resizebox{\hsize}{!}
    {\includegraphics[width=\hsize,clip]{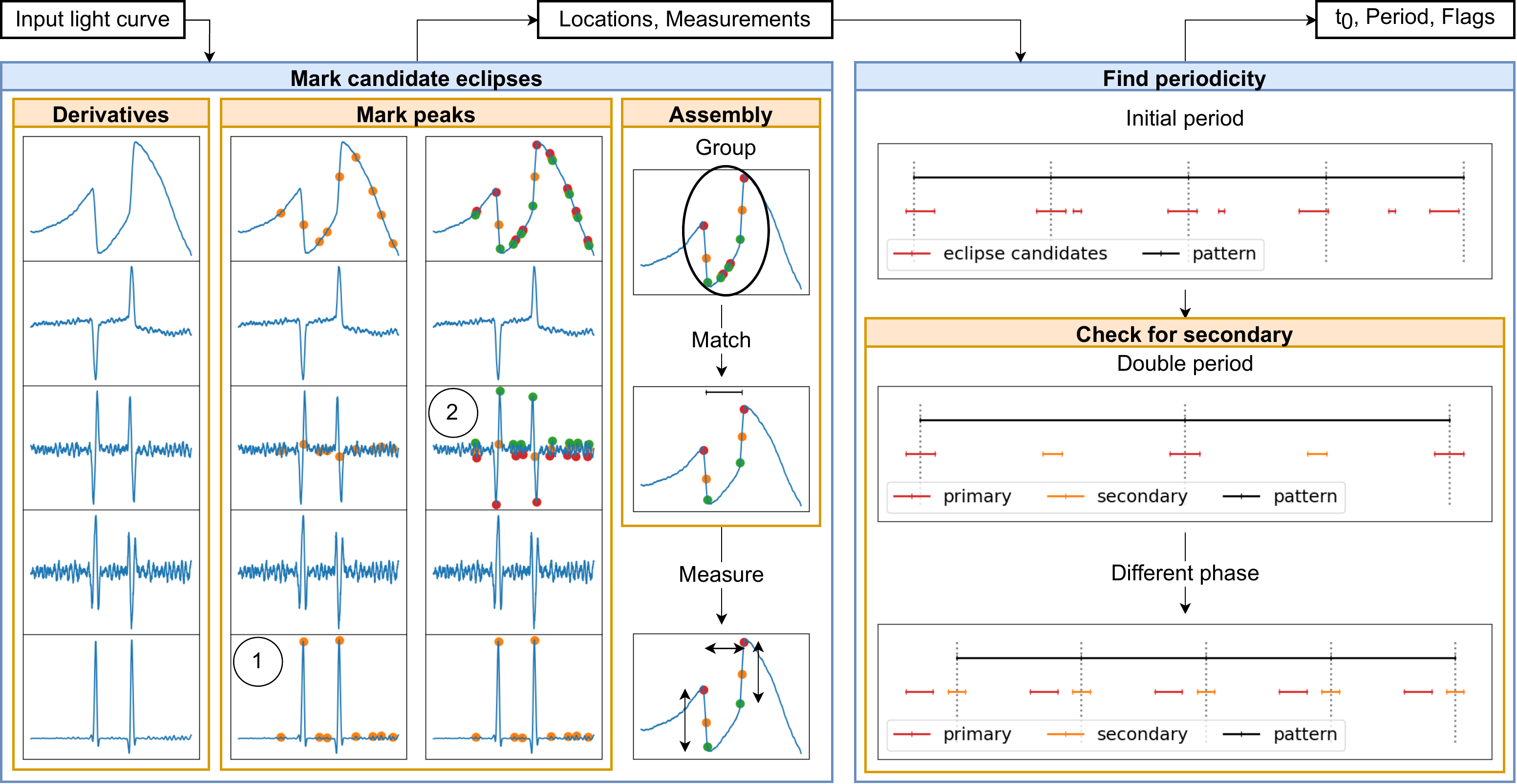}}
    \caption{Schematic overview of the main steps in the eclipse finding algorithm. It starts at taking time derivatives, marking peaks in those derivatives and assembling the marked points into eclipse candidates. These candidates are then matched to equally spaced patterns in order to find the most prominent periodicity and possibly discern between primary and secondary eclipses through differences in width, depth, SNR or location in orbital phase.}
    \label{fig:flow}
\end{figure*}

The resemblance between the first and third derivative presents the opportunity to increase the eclipse in/egress signal even more with respect to noise levels, for the purpose of algorithmically pinpointing these features. By multiplying these two derivatives, and taking the negative of that for convenience, we create a curve where noise peaks are largely smoothed out while the coherent signal produced by the presence of an eclipse is amplified (bottom panels in Fig.\,\ref{fig:derivs}). The features in this quantity correspond to positive peaks, so that an existing peak-finding algorithm can be used to locate them. The steps described here are schematically shown in Fig.\,\ref{fig:flow} as a flow chart of the main functions in \texttt{ECLIPSR}. Finding the peaks in derivative one-three is marked with an encircled 1.

From the sign of the third derivative at each peak location, we know if we have a possible eclipse ingress or an egress. While the first derivative could in principle be used for this purpose as well, the more pronounced presence of coherent intrinsic variability signal in the first derivative makes this an unreliable measurement. Moreover, the running average of the second and third derivative are much closer to zero than that of the first derivative.

The positions of the peaks in the second derivative are found by walking outward from the now known positions of the peaks in derivative one-three. This is marked in Fig.\,\ref{fig:flow} with an encircled 2. The peaks of the first derivative roughly correspond to the steepest part of the slopes of the eclipse. The second derivative has peaks where the first derivative is steepest: in other words, roughly where the light curve is changing direction most drastically. The positions of these peaks in the second derivative are good indicators for the start and end points of the eclipse, as well as marking the flat bottom of an eclipse.

At this stage, a crude SNR is measured for each candidate ingress and egress, and a cut is made based on this value to reduce the amount of spurious peaks that gets to the stage of assembling the full eclipses. The SNR measure is constructed by adding the individual SNR values calculated from each of the derivatives and the light curve itself. Noise levels for the light curve and the first time derivative are estimated by taking the average of the out-of-eclipse absolute point-to-point differences, while for the other three derivatives the noise is estimated by taking their average absolute value out-of-eclipse. The difference is made because the first two curves have significant non-zero running averages, while the latter three are close to zero on average when not in an eclipse. Note that `in eclipse' at this stage means `in ingress' or `in egress', because the full eclipses are not assembled yet.

The signal in the SNR measurement is obtained roughly in the same way for each curve, by measuring the height of the peaks above the background signal, or just the eclipse depth in case of the light curve, with the exception of derivative two. For the second derivative the signal is measured as the peak-to-peak difference between the negative and the positive peak that belong to the same ingress or egress.

\subsubsection{Treatment of noise in the derivatives}

Taking the time derivatives of a discrete light curve produces noise, which drowns out the eclipse signal. In all but the most noisy cases, this can be resolved by smoothing the curves before computing the derivatives to reduce short time-scale scatter while  increasing coherent signals. This is done by convolving the curve with a flat (constant) convolution kernel (i.e.\ a running average) with width $n_{kernel}$.
Gaps in the data are taken into account by repeating the data points at the edges several times before convolution, avoiding the merging of signal from two detached parts of the time series. The third derivative is also smoothed before multiplication with the first derivative, and the resulting curve (dubbed derivative one-three) is once again smoothed for optimal effect. 

The downside to this approach is that a new parameter is introduced ($n_{kernel}$). A wider kernel implies averaging over more data points, thus increasing the smoothing effect. Finding an appropriate value for the width of the smoothing kernel is not trivial. The implemented method relies on finding the eclipses in the light curve for a range of values. The best value for $n_{kernel}$ is picked by optimising the slopes, depths and signal-to-noise (SNR, see description below) values of the detected eclipses in each iteration, while incorporating two constraints in order not to smooth the light curve too much (or too little). The first constraint is based on a runs-test for non-randomness \citep{Bradley1968}, used on the original minus the smooth light curve: this quantifies the deviation of the smoothed curve from the original data by counting runs of consecutive points above or below zero. For values of $n_{kernel}$ where the test statistic is above one, the optimisation parameter is divided by the test statistic. The runs-test is also used to select only those values of $n_{kernel}$ for which a high (>100) maximum SNR value was found if there is no significant deviation from the original light curve for any value of $n_{kernel}$. This prevents under-smoothing in some cases. The second constraint uses derivative one-three by taking the squared mean of the absolute values and taking the reciprocal of one plus the result: 
\begin{equation*}
    \left(1 + mean\left(abs(derivative_{1x3})\right)^2 \right)^{-1}. 
\end{equation*}
This is closer to one when derivative one-three comes closer to zero: a maximum value for $n_{kernel}$ is set where it reaches 0.999.

\subsubsection{Assembly of the eclipse signal}

The detected peaks belong to separate eclipse ingress and egress candidates, or false detections. Due to the possibility of noise or other false peaks getting in between an eclipse ingress and egress, we cannot simply pair up consecutive candidates. To assemble the candidates into full eclipses, they are first binned into one, two, or three SNR groups and treated from highest to lowest SNR group. The groups are determined from the distribution of SNR values, which in the best cases can separate primary eclipses, secondary eclipses and noise peaks. Within each group they are then cut into sets of consecutive ingress peaks followed by consecutive egress peaks. For each set, the best match of ingress and egress is based on the measured depth, width and SNR of each in/egress; left-over in/egress peaks are transferred to the next SNR group. Making this division into SNR groups increases the success rate of the eclipse assembly.

At the end of the assembly, an array of indices is produced marking the eclipse start, left bottom, right bottom and end points for each eclipse as detected by the algorithm. Left and right bottom coincide when no flat bottom is detected\footnote{This does not exclude the presence of a flat bottom.}.

\subsection{Finding periodicity in the eclipse candidates}

The goal of the second part of the algorithm is to find the orbital period of the analysed EB, or half the orbital period if no distinction can be made between primary and secondary eclipses. Full eclipses are treated slightly differently than the eclipse candidates with only an ingress or an egress, which are referred to as half eclipses. Before looking for periodicity, some eclipse properties are computed. The midpoint is determined as the average between the times at each of the four marked eclipse indices. For half eclipses the bottom marker is taken to be the best estimate. Depths are measured for each side of the eclipse separately and averaged for full eclipses. Widths are measured between the furthest measured points of the eclipses, and a ratio is given between the width of the bottom of the eclipse and the full width as an indication of a flat bottom.
For optimal speed, further calculations are done with the per-eclipse measurements only and the time series is not used again until the very last part of the algorithm.

The function used to make an initial guess of the orbital period overlays an equally spaced pattern on the eclipse midpoints. Eclipses are said to be matching the pattern if they are the closest eclipse to a point in the pattern and within a quarter of the eclipse width of that point. The goodness-of-fit function for each period step is:
\begin{equation}
    {\rm g.o.f.}_{P} = \left(\frac{N_{\rm match}}{N_{\rm pattern}} \sum_{\rm matches} {\rm SNR}\right) - \sum_{\rm matches} d\;
    \label{eq:gof}
\end{equation}
where $d$ denotes the distance between an eclipse midpoint and the pattern point it was matched to (measured in days). The second term acts as a correction to the dominant left term which in itself cannot distinguish between very small changes in period, while the summed distances can.

A pre-determined range of periods is scanned and the highest goodness-of-fit determines the best period. This range depends on the found times of eclipse midpoint: the minimum period is determined by taking 0.95 times the minimum separation between eclipse candidates, the maximum period is three times the median separation between eclipse candidates.
It is then checked whether doubling the period results in secondary eclipses that are distinguishable from the primary eclipses. If this is not the case, it is checked whether there are eclipses that have the same period, but are at a different phase offset than the halfway point (for eccentric cases).

The outputs at this stage are the midpoint of the first (full) primary eclipse, the (orbital) period and a set of flags marking each eclipse candidate as either a primary (p), secondary (s) or other/possible tertiary (t). The third category includes any candidates that do not match the pattern of the primary or secondary eclipses, so in principle eclipses of a third body could be in this list along with noise peaks that are picked up.

\begin{figure*}
\resizebox{\hsize}{!}
    {\includegraphics[width=\hsize,clip]{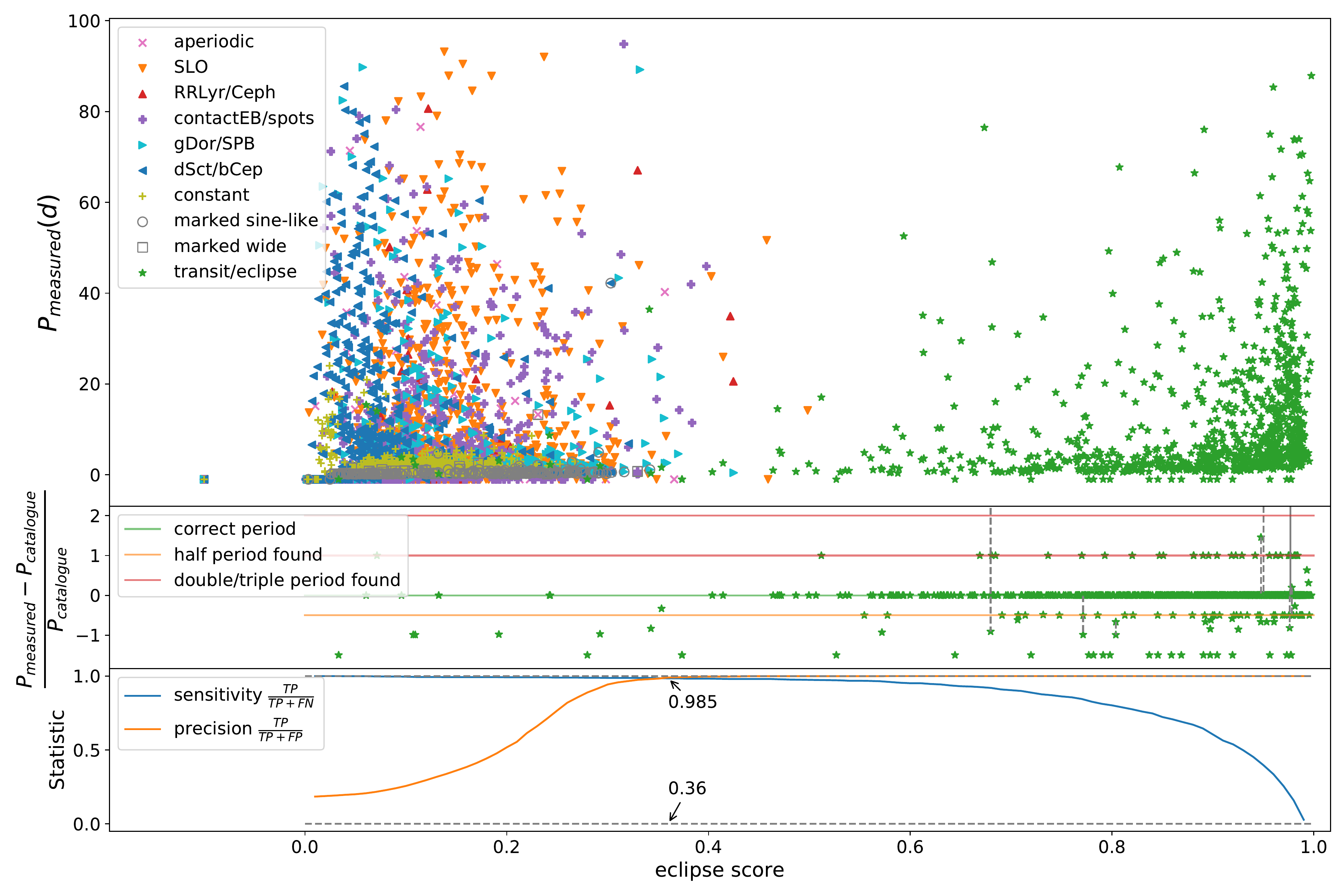}}
    \caption{Results of the \textit{Kepler} test set as a function of eclipse score: (top) the measured periods for all colour coded categories in the set, (middle) the accuracy in the measured periods for the eclipse category as compared to the \textit{Kepler} EB database, (bottom) the sensitivity and the precision with arrows indicating the ordinate and abscis values of where the two curves intersect.}
    \label{fig:testset}
\end{figure*}

\subsection{Assigning an eclipse score}

The final stage of this algorithm is to assign a value to each analysed light curve to give an indication of the likelihood that it indeed contains eclipse-like signal. Five quantities between zero and one are computed and combined in a formula to end up with a single value that is closer to one for light curves containing eclipses and closer to zero for other categories of targets. The formula for the eclipse score is
\begin{equation}
    {\rm score} = A_0 \cdot A_2 \cdot \frac{\sqrt{A_1^2 + A_2^2 + A_3^2 + A_4^2}}{2}, 
    \label{eq:score}
\end{equation}
with explanations of the five quantities ($A_i$) as follows. 

\begin{itemize}
\item 
$A_0$ simply transforms the sum of the average SNR of the primary eclipses and the average SNR of the secondary eclipses to a value between zero and one.

\item
$A_1$ is the ratio between the number of found primary and secondary eclipses and the total number of possible eclipses in the time span of the light curve given the ephemeris and taking into account gaps in the data.

\item
$A_2$ measures the average slope of the eclipses divided by the median of the absolute first time derivative of the light curve outside of the eclipses. This value is transformed analogously to $A_0$ to a domain between zero and one.

\item
$A_3$ is a measure for how symmetric the eclipses are, by looking at the depths and slopes on the left and right side of the eclipses. The difference in depth measured at eclipse ingress and at egress is divided by their sum and the same is done for the slope at these locations. This is averaged over all eclipses and transformed to a value between zero and one.

\item
$A_4$ is a measure for how similar the eclipses are to each other (handling primary and secondary eclipses separately), by taking the mean of the squared and mean-normalised differences from the mean for the SNR, depth, and width values and transforming them to a combined value between zero and one.
\end{itemize}
Expressions for $A_i; i=0, \ldots,4$ are provided in Appendix\,\ref{apx:formulae}, as well as some example light curves and their values for each of these quantities.

\begin{table}
	\centering
	\caption{Description of the test set.}
	\label{tab:trainig_set}
	\begin{tabular}{p{2.5cm} p{4.5cm} l}
	\hline
	Category label & Type  & Size \\
	\hline
	aperiodic & Aperiodic stars & 830\\
	contactEB/spots & Contact binaries and rotational variables & 2\,260 \\
	dSct/bCep & $\delta$ Sct and $\beta$ Cep stars & 772\\
	transit/eclipse & Eclipsing binaries & 974\\
	gDor/SPB & $\gamma$ Doradus and SPB stars & 630\\
	RRLyr/Ceph & RR Lyraes and Cepheids & 62\\
	SLO & Solar-like pulsators & 1\,800\\
	constant & Constant stars & 1\,000 \\
	\hline
	\end{tabular}
	\tablefoot{This Table is reproduced from \citet{Audenaert2021}.}
\end{table}

An advantage of this method is that it looks at the local slope in a light curve to determine the occurrence of an eclipse, so it is only locally affected by the presence of other variability. This (intrinsic) variability can be of arbitrary amplitude as long as the eclipses reveal themselves with respect to the noise in the data. That leads us into a disadvantage, which is that each eclipse has to be detected separately and the signal is not boosted by the presence of other eclipses (e.g. as in the BLS method). This means low SNR eclipses are hard to find with this method, the advantage being that it has the potential to find singly eclipsing systems.

\section{Application to a \textit{Kepler} test set}
\label{sec:testing}

To evaluate the performance of the described \texttt{ECLIPSR} algorithm, a labelled test set consisting of 6423 \textit{Kepler} targets is used, whose light curves were restricted to 27.4\,d to mimic TESS data. The \textit{Kepler} EB Catalog (third revision) \citep{2011AJ....141...83P, 2011AJ....142..160S, 2016AJ....151...68K} is used to obtain periods for each of the targets in the transit/eclipse category. The test set is described in detail in \citet{Audenaert2021} and is freely accessible through the TESS Asteroseismic Consortium (TASOC) Wiki pages\footnote{\url{https://tasoc.dk/tda/TrainingSets}}. The set is summarised in Table\,\ref{tab:trainig_set} and contains light curves of a variety of stellar variability classes including radial and non-radial pulsators, rotational variables and contact binaries. The set also includes synthetic light curves mimicking constant stars with noise following the performance of the \textit{Kepler} space telescope. This test set made a rough distribution of all sorts of variables in a limited number of dominant variability categories and is therefore limited in its representation of the nature of the variables. In other words, it has an artificially enforced distribution of targets between too few categories compared to the complex variability tree defined by \citet{Eyer2019}. In particular, the ratio of eclipsing to non-eclipsing systems in this \textit{Kepler} test set is higher than in a randomly chosen \textit{Kepler} data set. Nevertheless, the test set was specifically designed to prepare optimal variability classification for TESS data and hence it is also suitable to test our \texttt{ECLIPSR} algorithm on the aspect of EB variability.

\begin{figure*}
\resizebox{\hsize}{!}
    {\includegraphics[width=\hsize,clip]{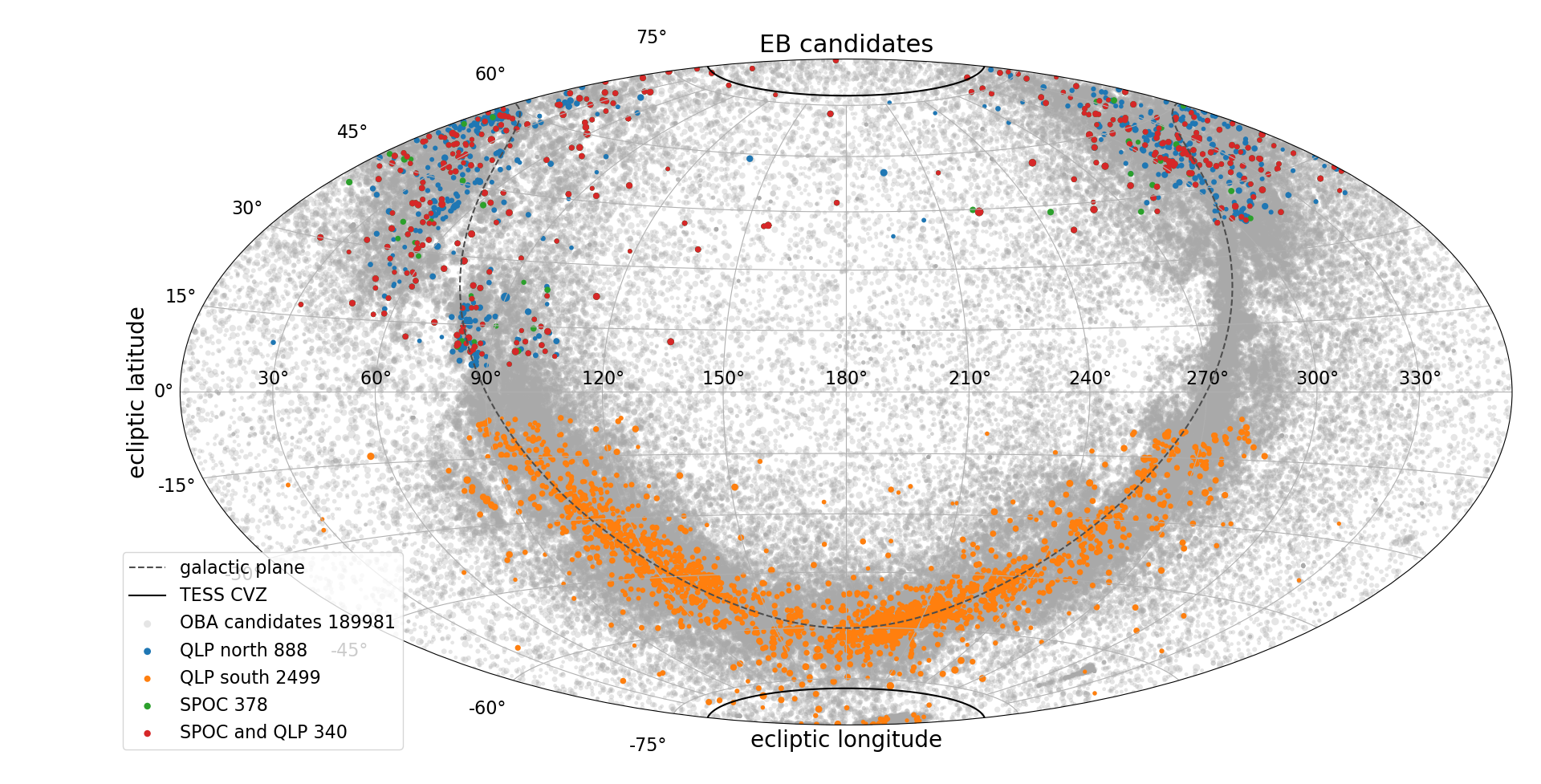}}
    \caption{Distribution on the sky of the overall sample (grey) and the EB candidates (other colours) in ecliptic coordinates. The TESS CVZs are indicated with thick solid lines and the galactic plane is indicated by the dashed line.}
    \label{fig:ecliptic}
\end{figure*}

Figure\,\ref{fig:testset} shows the output of \texttt{ECLIPSR} for the whole test set, revealing several aspects. The top panel shows the period that was found by the algorithm for each target as a function of the eclipse score; each category of the test set has its own symbol and colour. Light curves that have a zero period and negative score had no eclipse detections.
The middle panel of Fig.\,\ref{fig:testset} shows how close the measured period is to the one listed in the \textit{Kepler} EB Catalog. The horizontal lines indicate whether a certain multiple of the catalogue value was found: zero indicates the correct period, minus one half means half the period was found and plus one means twice the period was found. A value of -1.5 in this plot is artificially inserted and means no period was found for that target. In practice this means that one or more eclipses were missed by the algorithm. The high eclipse score of these targets implies that even single eclipses can in principle be identified by our method, unlike for Fourier-based methods. Two examples are given in Fig.\,\ref{fig:lcs-special} (TIC 385663527 and TIC 130415266).

The bottom panel of Fig.\,\ref{fig:testset} shows the curves of sensitivity and precision as functions of the eclipse score. If a cut is made in eclipse score, keeping only the targets above that value, the sensitivity is the fraction of EBs that are left from the total number of EBs and the precision is the fraction of the selected targets that are EBs. In terms of true (T) and false (F) positives (P) and negatives (N), these statistics are expressed as:
\begin{align}
\text{sensitivity} = \frac{\text{TP}}{\text{TP + FN}}, && 
\text{precision} = \frac{\text{TP}}{\text{TP + FP}}.
\end{align}

To find a balance between these two statistics and make an informed cut in the eclipse score, one can use the point at which the two curves intersect. This way, we maximise how many of the EBs are filtered out of the total set of targets, while minimising the amount of non-EBs that get through the selection. The higher the value of the curves at this intersection, the better the algorithm performance.
What this means in practice for the presented algorithm is that if a cut is made at an eclipse score of 0.36, about 98\% of all EBs in the test set are retrieved and about 2\% of the selected targets are false positives (see the arrows indicating the intersection in the bottom panel of Fig.\,\ref{fig:testset}). This is not necessarily a good representation of the real world performance, as this test set has an artificial ratio of eclipsing to non-eclipsing systems that is much higher than in a randomly chosen real world data set. 

\section{Application to \textit{TESS}}

As explained in Section\,\ref{selection}, the selection criteria for intermediate- to high-mass candidates resulted in 189\,981 targets for which the (available) light curves are examined for eclipses. Cross-matching with the publicly released TESS-SPOC 30-min cadence light curves results in 14970 targets with light curves of various lengths. 91142 of our targets have a light curve in the MIT QLP data. We applied the described \texttt{ECLIPSR} algorithm (version 1.0.2) to these light curves and used the cutoff at the eclipse score value of 0.36. This resulted in a set of 457 EB candidates for the TESS-SPOC data and 5418 EB candidates for the MIT QLP, totalling 5502 unique targets. 

These candidates were subsequently manually inspected to exclude clear false positives. Apart from a number of light curves with artefacts that were wrongly identified as eclipses, a notable class of contaminants appears to be stars with very sinusoidal light curve patterns, showing two dominant frequency peaks in their Lomb-Scargle periodograms. This sub-group of targets with the most sinusoidal light curves could be contact systems showing the O'Connell effect \citep{1968AJ.....73..708M} or ellipsoidal variability, but are indistinguishable from for example single-mode pulsators or rotational modulation signal, and have been excluded from the EB candidates.

After visual inspection and elimination of the false positives, the TESS-SPOC sample has 378 remaining targets and the MIT QLP sample has 3387, which means the success rate on these particular TESS data sets is about 83\% and 63\% respectively. The lower success rate can largely be attributed to artefacts of the data reduction that cause high peaks in the derivatives, which can lead to a high eclipse score if the artefact passes the checks for being sufficiently eclipse-like. In total, 3425 unique targets exhibit eclipse signal. These remaining systems cover a wide range in light curve morphology. The \texttt{ECLIPSR} scores favour the steep eclipses and sharp transitions from detached systems. However, this does not mean that contact systems are all eliminated. During inspection, 207 likely cases were found and flagged as possible contact systems.

As was shown to be possible by the tests, 227 singly eclipsing systems were also picked up by \texttt{ECLIPSR} (0.2\% of all the analysed light curves). While the direct benefits for light curve modelling are small, these targets are not excluded from the sample to allow for follow-up observations. These targets have an assigned period of -1, and t$_0$ indicates when the single eclipse (or the primary eclipse if a clear secondary is present) occurs. The smallest eclipse depth found this way is half a percent for a handful of cases, increasing to a dozen cases at the percent level. The noise level for all of these is lower than the eclipse depth, with the minimum depth to white noise scatter ratio being around two to three (see example TIC 130415266 in Fig.\,\ref{fig:lcs-special}). 

\subsection{Flux contamination}

An unavoidable source of contamination comes from flux of neighbouring sources seeping into the photometric aperture used for light curve extraction. This is due to a combination of the large 21 arcsecond pixel scale of the TESS CCDs and high densities of sources. Determining with certainty whether eclipse signal is coming from the intended target or a nearby source is not trivial and is beyond the scope of this work, as it requires independent data. A helpful indicator for contamination is the level of flux in the aperture due to sources other than the intended target. Such a measure is given in the TESS-SPOC data, in the form of the \texttt{CROWDSAP} parameter, which gives the fraction of flux from the target divided by the total flux in the aperture \citep{2021ApJS..254...39G}. 

An indication that contamination occurs for some of the candidate EBs is given by the fact that many of them have visually identical eclipse signals to their closest sky-projected neighbours. To investigate this further, we grouped candidates based on their periods and angular separation. Using a relative period tolerance of 1\% and one degree as the maximum angular separation, we find 207 groups containing a total of 441 candidates. This excludes 37 cases where the eclipse signal was actually not the same, but just happened to be close in period and angular distance. The groups are numbered and these numbers are included in the EB candidate catalogue. Further investigation is needed to point out which of these are the genuine EBs. Adjusting the number of unique EB candidates for these duplicates results in 3155 remaining candidates. That is not taking into account flux contamination of sources outside of our EB candidate sample.

\subsection{Spatial and orbital period distributions}

Figure\,\ref{fig:ecliptic} shows the distribution of targets on the sky in ecliptic coordinates. Figures with galactic and equatorial coordinates are available in Appendix \ref{apx:sky-dist}. In grey is the overall sample selected from the TIC, which are independent from the TESS observations and thus cover the whole sky. The galactic plane is clearly visible as a strongly increased density of sources, and the EB candidates naturally follow this same density distribution. Since the observing sectors of the nominal TESS mission do not cover each and every square degree, most notably along the ecliptic plane, there are gaps where no EB candidates are present. The northern-CVZ contains only three candidates, while the southern-CVZ contains 45 (excluding three duplicates), 34 of which in the area covered by the the Large Magellanic Cloud (LMC). One target in the northern-CVZ is a known EB (TIC 377192659) and 14 of the targets in the southern-CVZ are known EBs in the literature. TIC 31265416 is a confirmed Galactic foreground star to the LMC, while some 20 targets are reported to be part of the LMC.

\begin{figure}
\centering
\includegraphics[width=\hsize]{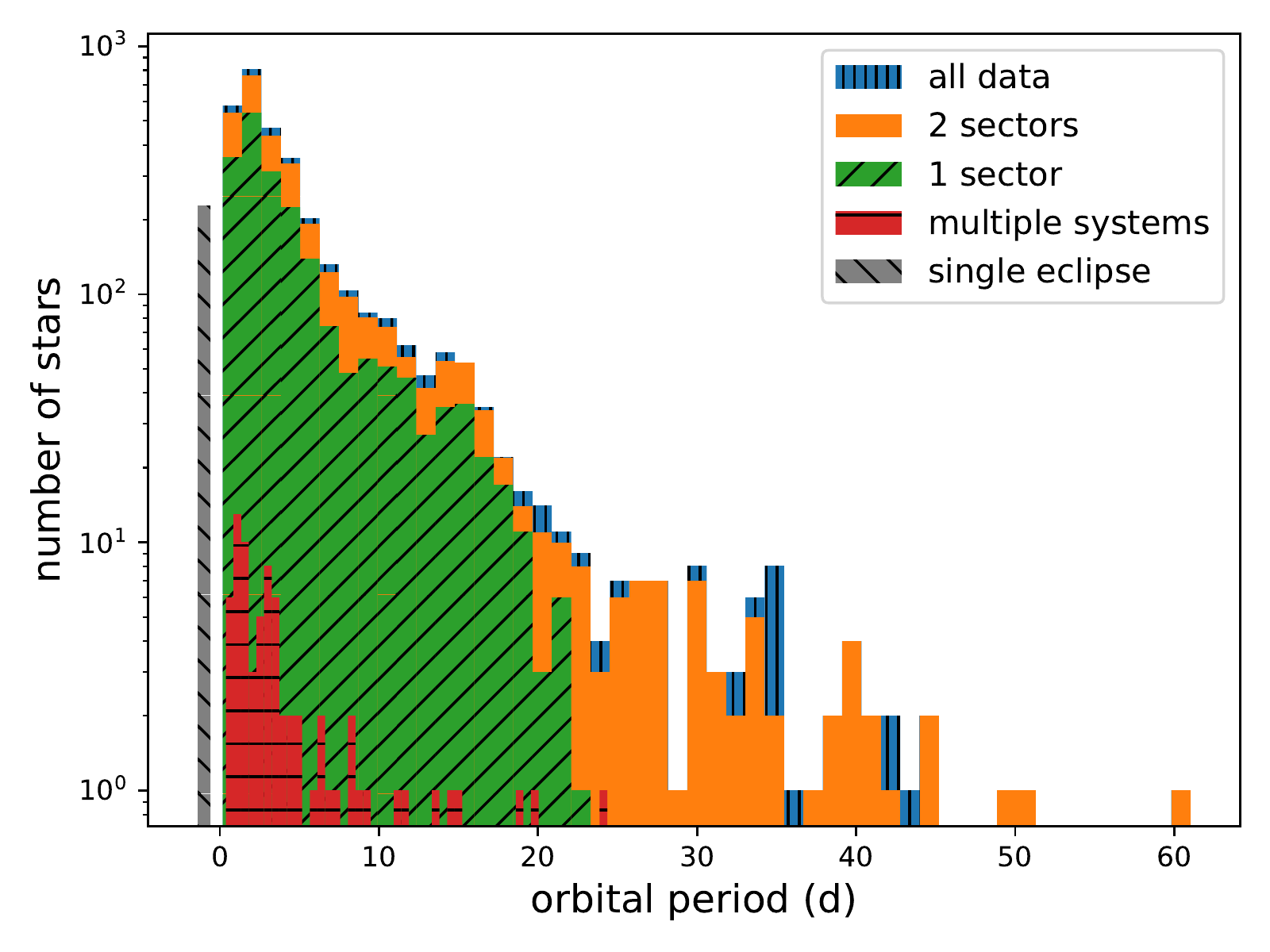}
  \caption{Period distribution below 61\,d of all EB candidates after removal of false positives and correction of the periods where necessary. Green and orange bars show the distributions of targets with only one or two sectors available, respectively. The red bars indicate the periods of eclipse signals that are present in addition to the one marked by \texttt{ECLIPSR}, pointing at triple or quadruple systems.}
     \label{fig:period-dist}
\end{figure}

The distribution of orbital periods below 61\,d is shown in Fig.\,\ref{fig:period-dist}. Four systems have a longer period and are not shown in the histogram. Green bars show the distributions of targets with only one sector available, orange bars add to that the targets with two sectors and blue shows the distribution for all targets. Note that these sectors are not necessarily consecutive, and that the log-scale means that height difference is not representative of number. In ambiguous cases where no secondary eclipses are distinguished, the period is set to the lowest possible given the found eclipses. In case the primary and secondary eclipses actually have the same shape, the orbital period is twice the reported period. These cases are to be confirmed by radial velocity measurements. The thinner red bars in Fig.\,\ref{fig:period-dist} indicate periods of possible triple or quadruple systems. Not all possible multiple (\textgreater 2) systems identified have enough eclipses in the light curve to have a reported period. As expected for these intermediate- to high-mass systems, the period distribution favours short periods \citep{2012Sci...337..444S, 2014ApJS..213...34K, 2015A&A...580A..93D, 2017A&A...598A..84A}. 
The distribution approaches zero as the orbital periods get longer and approach  the one-sector timescale of 27\,d because it becomes more likely that only one eclipse is covered by the data points. For orbital periods above 27\,d a single sector of data could easily miss an eclipse altogether. A total of 63\% of the light curves in our sample are only a single sector in length, followed by 31\% having two sectors of coverage and the remaining 6\% has at least three sectors.

\begin{figure}
\centering
\includegraphics[width=\hsize]{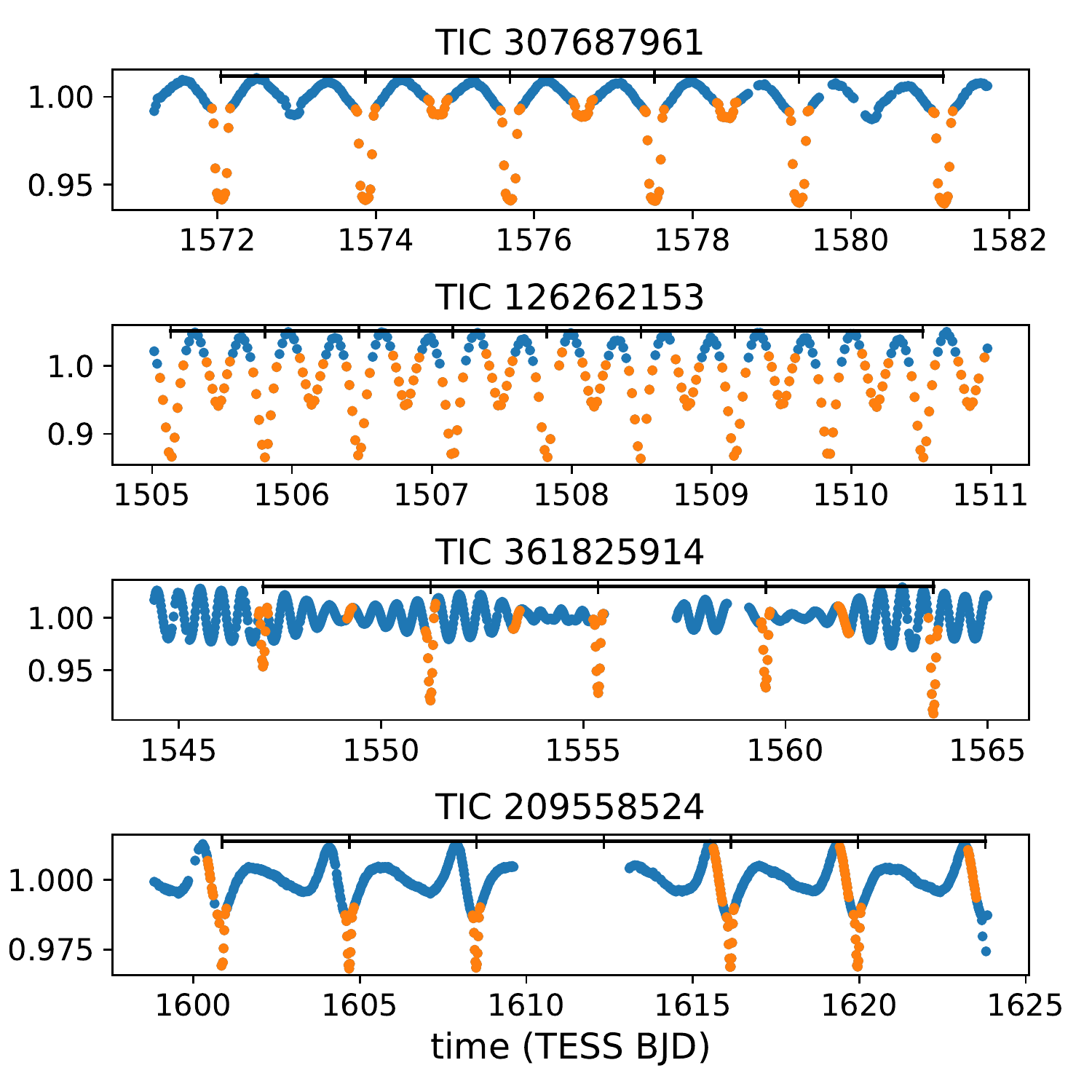}
  \caption{Part of the TESS light curves of four example EB candidates in different categories. Top to bottom: a case with a clear secondary and flat-bottomed eclipses, a candidate contact system, an example with likely pulsations and a heartbeat system. The orange points indicate the candidate eclipses as marked by \texttt{ECLIPSR}. Full and folded light curves for these targets can be found in Figures\,\ref{fig:lcs}, \ref{fig:lcs-puls} and \ref{fig:lcs-special}.}
     \label{fig:ex-short}
\end{figure}

Systems where a secondary eclipse was identified are tagged in the catalogue. An example with flat-bottomed eclipses is shown in Fig.\,\ref{fig:ex-short} (TIC 307687961). Tags are also used to indicate the presence of other types of variability, such as intrinsic variability, candidate contact systems, or heartbeat-like light curves. Most of the heartbeat-like light curves found also show clear eclipses (37 of 40), like the example in Fig.\,\ref{fig:ex-short} (TIC 209558524). In fact, only three cases were found with ambiguous eclipse detections (TIC 261617730 - shown in Fig.\,\ref{fig:lcs-special}, TIC 295158448, TIC 173301640). The category of EB candidates with intrinsic variability contains 1008 targets, of which more than three quarters are likely pulsating. This is a large fraction of the total EB sample, which is promising for the combined modelling of eclipse and pulsational signal. 188 candidate contact systems are marked.

A range of light curves is shown in Figures \ref{fig:lcs}, \ref{fig:lcs-puls}, \ref{fig:lcs-var} and \ref{fig:lcs-special} as examples. Descriptions and discussion of some of the examples are provided in Appendix \ref{apx:examples} as well.

\begin{figure}
\centering
\includegraphics[width=\hsize]{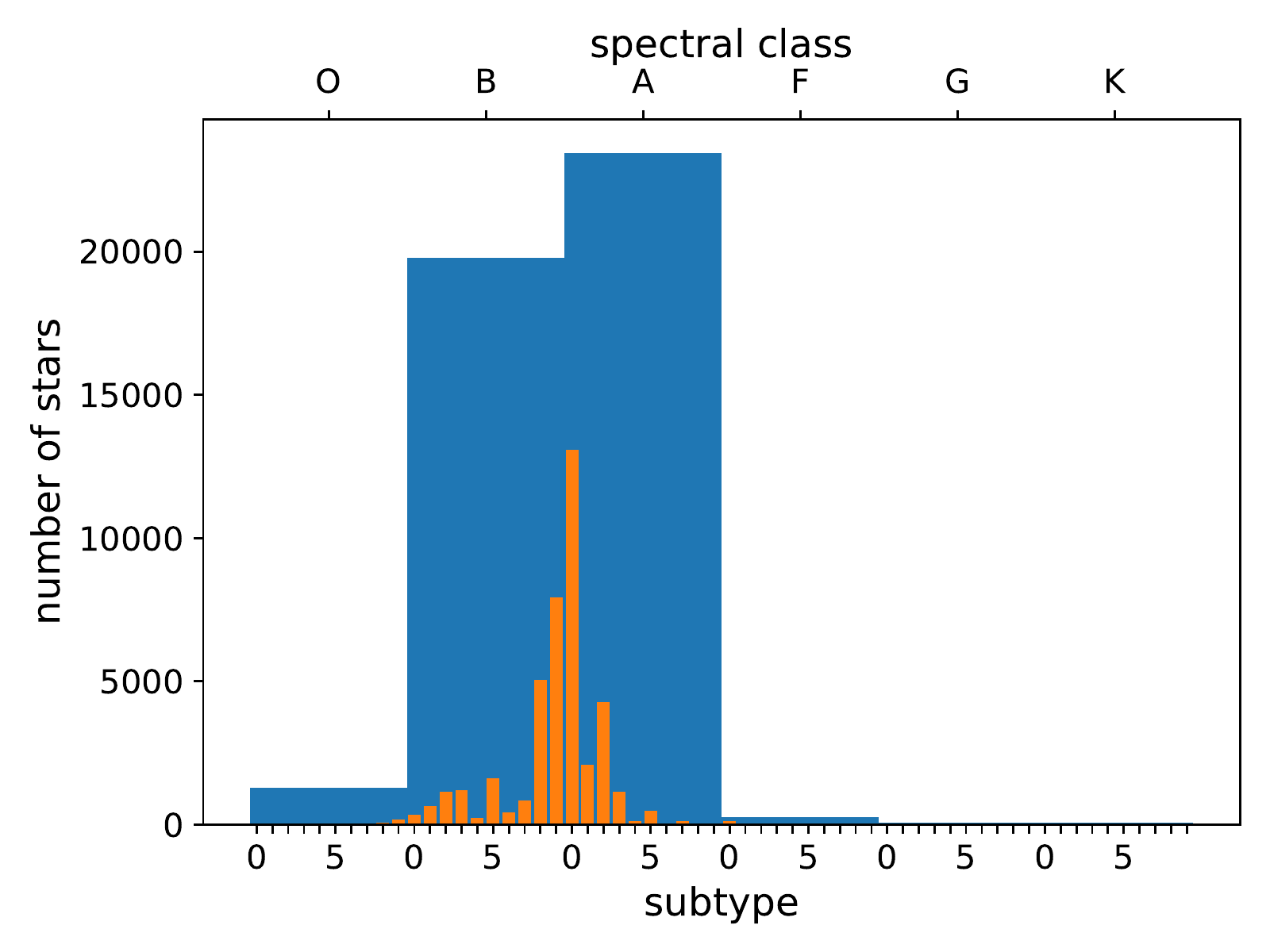}
  \caption{Spectral classes and subtypes for 45142 targets in the overall catalogue, obtained from \texttt{SIMBAD}. The subtype numbers do not add up to the amounts of the spectral classes since part of the spectral types only contain the class letter.}
     \label{fig:spectypes}
\end{figure}

\section{Cross-match with published catalogues}
\label{cross_match}

To gain insight in the distribution of spectral types of the stars in our overall sample, we obtained the available spectral types from \texttt{SIMBAD}. In total, a spectral type was found for 45142 targets, of the 85083 targets from the overall sample (with and without light curves) that returned a query result from \texttt{SIMBAD}. These are grouped according to their spectral class and integer subtype and the group counts are displayed in Fig.\,\ref{fig:spectypes}. Only the first part of each \texttt{SIMBAD} spectral type is used, so type names with for example "B3/4" are counted as B3. The numbers in the subtype bins do not add up to the total amount in the respective spectral classes, because a significant number of targets do not have a subtype specified. 

We plot the distribution on the sky of targets with a reported spectral type on a backdrop of the total intermediate- to high-mass catalogue (see Fig.\,\ref{fig:simbad-sky}). The targets with reported spectral type generally follow the overall density pattern of sources, indicating no intrinsic bias toward any specific spectral type from a spatial argument. In conjunction with the sufficiently large number of targets with reported spectral type, covering $\sim$24\% of the catalogue, we take this as a good representation of the overall sample. This means the contribution of A-type stars amounts to $\sim$52\%, with even cooler stars (F/G/K-types) at $\sim$0.9\%. The spectral (sub)type with the largest number of targets is A0, contributing $\sim$32\% to the total. O-type stars amount to $\sim$2.9\% and the hottest reported sub-type is O2 (3 cases).

\begin{table}
	\centering
	\caption{Result from the cross-matches with other catalogues.}
	\label{tab:catalogues}
	\begin{tabular}{p{2.2cm} p{1.5cm} p{1.5cm} l}
	\hline
	catalogue & in OBA catalogue & with light curve & in EB candidates \\
	\hline
	\texttt{SIMBAD} "EB*" & 684 & 430 & 333\\
	\texttt{SIMBAD} "**" & 2\,238 & 1\,715 & 88\\
	\texttt{SB9} & 417 & 299 & 93 \\
	\textit{Kepler} EB & 40 & 36 & 10\\
	\hline
	\end{tabular}
	\tablefoot{Result from the cross-matches with \texttt{SIMBAD}, the \texttt{SB9} catalogue and the \textit{Kepler} EB Catalog. See text for an explanation of the columns.}
\end{table}

To get a sense of how many known binaries were found and missed, \texttt{SIMBAD} was also queried for object type and the EB candidates cross-matched  with the \texttt{SB9} \citep{2004A&A...424..727P} catalogue and the \textit{Kepler} EB Catalog.
The results are summarised in Table\,\ref{tab:catalogues} in terms of the amount of overlap in three categories (corresponding to the columns). The first number gives the amount of targets that also occur in the overall catalogue presented in this work. The second number shows how many are left over if we only look at the targets for which we have light curves in the TESS-SPOC and/or MIT QLP data. The last column contains the amount of overlap with our EB candidate catalogue. There are two rows for \texttt{SIMBAD}, which are for two different selected object types: double or multiple stars ("**") and eclipsing binaries ("EB*"). 

The number of known EBs (the "EB*" object type and the \textit{Kepler} EB Catalog) can give an idea of the number of EBs that were missed by our eclipse search in the examined TESS data sets. The number in the second to last column gives the sum of binaries that were identified plus the targets that were missed. Taken at face value, if we divide the 333 EB candidates with the label "EB*" by the 430 targets with that label that have a light curve in our data sets, we arrive at a true positive rate (i.e. sensitivity) of 0.77 to find eclipsing systems.

When examined more closely, we find that the 36 targets that were missed in the \textit{Kepler} EB Catalog fall into roughly four categories. Most of the light curves (14 cases) are very sinusoidal in shape, which our method is less sensitive to, due to the reliance on peaks in the derivatives of the light curve. A second large part of the missed EBs (8 cases) is light curves exhibiting variability other than the EB signal, to such a degree that the eclipses are not picked up by our method, or generally odd shaped light curves that are not expected to be successfully recovered. Another possibility is that the eclipses are simply not present in the relatively short available time series (3 cases). The leftover category are genuine false negatives in the sense that the \texttt{ECLIPSR} algorithm should have found them: there is only one such case in the missed \textit{Kepler} EB Catalog targets, which has an eclipse score of 0.326, under the threshold of 0.36. This extends to the \texttt{SIMBAD} "EB*" category, with 7 (of 127) cases that should have passed, having eclipse scores between 0.26 and 0.34.

Similarly to the \textit{Kepler} EBs, we can use the ephemerides of the SB9 catalogue to look for EB signal that was missed. In this set of 206 missed binary systems, 43 sinusoidal light curves are found. A set of 79 light curves either fall into the category of small relative eclipse signal or odd shapes, or have no visible eclipses at the given ephemeris\footnote{The SB9 catalogue is not an EB catalogue after all.}. For 81 targets there was insufficient coverage to be able to find eclipses at the given ephemerides. The leftover 3 targets are actual false negatives with 0.341, 0.259 and 0.218 for their eclipse score. 

\section{Conclusions}

We presented an all-sky catalogue of 189\,981 intermediate- to high-mass OBA-type dwarf candidates selected from the TIC, which was obtained by selecting stars brighter than 15 TESS magnitude and applying several filters to arrive at a sample that is not biased against binary systems. We selected for the temperature domain covered by intermediate- to high-mass stars by using colour cuts in the three 2MASS photometric bands (J, H and K) based on an existing sample of OB-type stars selected by spectral type. The resulting contribution of A-type stars is 52\%, the majority of which (32\% of the total) has spectral type A0. This catalogue was made as an intermediate step towards a sample of intermediate- to high-mass EB candidates with and without intrinsic variability observed by TESS.

In a recent work by \citet{2021A&A...650A.112Z} an OBA-type catalogue was produced using a Gaia magnitude cutoff of $G<16$ for the purpose of mapping hot stars in the Galaxy. Their catalogue contains 988\,202 entries with an estimated 45\% of the stars hotter than A0. This leaves our overall sample (with and without light curves) smaller by a factor five, although our sample has a slightly higher fraction of stars hotter than A0 (3\%). Most of the difference is due to the fainter magnitude cut of $G<16$, which corresponds to a TESS magnitude between roughly 15 and 16 for typical Gaia colours ($G_{BP} - G_{RP}$) of our stars \citep{2019AJ....158..138S}. A TESS magnitude cut at $T=15.5$ results in a factor two in volume reached compared to our cut at $T=15$. The colour cuts made are not easily compared, but we suspect that they can account for the remaining difference. In any case these two catalogues share sufficiently large overlap of OBA stars given their different selection criteria and purposes. \citet{2021A&A...650A.112Z} estimate their completeness level for O- and B-type stars to be around 86-90\% from cross-matching with the catalogues by \citet{2019ApJS..241...32L} and the Galactic O-Star Spectroscopic Survey \citep{2014ApJS..211...10S, 2016ApJS..224....4M}.

Light curves of the public data releases of the TESS-SPOC and MIT QLP of the TESS nominal mission were used in the selection of our OBA-type EB candidates. The light curves for a total of 91193 targets from the overall catalogue were searched for eclipses and an initial total of 5502 EB candidates were marked. 
The initial search for eclipses was done automatically, using a newly developed method as implemented in the \texttt{ECLIPSR} code, that discriminates as little as possible against EBs with other variability signals intrinsic to the star. This approach was taken in order to get a sample of EBs that both contain pulsators and non-pulsators in balanced amounts. After automatic selection a manual check was done on the EB candidates that passed the selection criterion of the automated code. During this manual check it became clear that stars showing particularly coherent intrinsic variability can be misidentified as EBs by this method, and that data processing artefacts in light curves are also a substantial source of contamination. We demonstrate that our method works sufficiently well for the purpose of selecting pure EBs as well as EBs whose individual components are intrinsically variable (Figures \ref{fig:lcs-puls} and \ref{fig:lcs-var}). The total number of EB candidates in the catalogue after manually removing false positives is 3425. Broken down by data set, there are 378 EB candidates in the TESS-SPOC light curves (initially 457) and there are 3387 EB candidates in the MIT QLP light curves (initially 5418). This shows that a larger fraction of the MIT QLP candidates were misidentified: 32\% versus 12\% for the TESS-SPOC data. This difference can be attributed to the light curves with a certain type of data processing artefacts that our eclipse search marked as possible eclipses, which are present in the MIT QLP data and not in the TESS-SPOC data

Our OBA-type EB candidate catalogue contains several pieces of information in addition to the ephemerides for each target. The average eclipse width and depth are provided, and the ratio between the width of the bottom of the eclipse and the total width. Furthermore we provide some tags that resulted from the manual inspection, including the presence of secondary eclipses, possible other variability, possible contact systems and possible heartbeat systems. The variability tag is subject to the same possibility of contamination as the eclipse signal itself, hence the use of the term `possible'.

A part of the binary and partially asteroseismic sample presented in this work will be further analysed in terms of binarity and pulsational signal in the light curves. This has the purpose to gather supplementary spectroscopy with the aim of stellar modelling, in order to improve our understanding of the rotation and mixing processes at work in the interiors of these hot stars. Such modelling will rely on the frequencies of detected and identified pulsation modes. In order to unravel these from instrumental frequencies with certainty, it is required to construct TESS light curves from customized masks per star instead of relying on the light curves from the public general pipelines as done here \citep{Tkachenko2013, Papics2014, Papics2015}. Moreover, the
focus will be on the targets with sufficiently long light curves to ensure sufficiently high precision in the pulsation frequencies \citep{Aerts2018}. Selection of the most appropriate subsample for asteroseismology will again be done in a maximally automated way to tackle the large number of targets in our catalogue that are now available for analysis, resulting in orbital parameters, masses, radii and asteroseismic quantities such as mode frequencies and their spacing patterns in the frequency or period domain \citep{Aerts2021}.

We plan to make a quantitative comparison between our method for eclipse detection and methods based on different mathematical principles in future work. The comparison between our results and those by \citet{Audenaert2021}, who used a machine learning approach to classify variable stars, reveals similar capacity. Our test in Section \ref{sec:testing} makes use of the same set of light curves and we find that we recover a similar fraction of EBs from that set at about 98\% compared to their 96\%.

Gaia DR3, arriving in 2022, will bring parallaxes and multiple radial velocity epochs to binary systems \citep{2016A&A...595A...1G, 2018A&A...616A...1G}, allowing the target selection to be expanded towards the lower mass stars that still have convective cores in the (late) A- and F-type regime by introducing an absolute magnitude cut. This concerns a much larger total number of stars, which emphasises the need for automated methods. Adding an absolute magnitude cutoff for these targets is important, because the region in the colour-magnitude diagram directly below this lower temperature regime is much more crowded and thus will result in more contamination from lower mass stars than was the case for the higher temperature regime treated here.

\begin{acknowledgements}

The research leading to these results has received funding from the European Research Council (ERC) under the European Union’s Horizon 2020 research and innovation programme (grant agreement N$^\circ$670519: MAMSIE), from the KU\,Leuven Research Council (grant C16/18/005: PARADISE), from the Research Foundation Flanders (FWO) under grant agreements G0H5416N (ERC Runner Up Project to AT), G0A2917N (BlackGEM), 1124321N (Aspirant Fellowship to LIJ), and 12ZB620N (Junior Postdoctoral Fellowship to TVR), and from the BELgian federal Science Policy Office (BELSPO) through PRODEX grant PLATO.
This paper includes data collected by the TESS mission, which are publicly available from the Mikulski Archive for Space Telescopes (MAST). Funding for the TESS mission is provided by NASA’s Science Mission directorate. We acknowledge the use of TESS High Level Science Products (HLSP) produced by the TESS Science Processing Operations Center (SPOC) (\href{https://dx.doi.org/10.17909/t9-wpz1-8s54}{DOI: 10.17909/t9-wpz1-8s54}) and the Quick-Look Pipeline (QLP) at the TESS Science Office at MIT (\href{https://dx.doi.org/10.17909/t9-r086-e880}{DOI: 10.17909/t9-r086-e880}).
We thank Dominic Bowman for useful discussions and comments on the manuscript.
This work and the presented code (\texttt{ECLIPSR}) make use of Python (Python Software Foundation. Python Language Reference, version 3.7. Available at \href{http://www.python.org}{www.python.org}) and the Python packages Numpy \citep{numpy}, Numba \citep{numba}, Scipy \citep{scipy} and Matplotlib \citep{matplotlib}.   
\end{acknowledgements}

%
%

\bibliographystyle{aa}
\bibliography{references}

\begin{appendix}
\section{Eclipse score supplement}
\label{apx:formulae}

The formulae for the five quantities going into the eclipse score (Formula\,\ref{eq:score}) are given here. We also provide some example light curves with the values for each quantity in Table\,\ref{tab:qtt-values}.

As stated in the main text, $A_0$ simply transforms the sum of the average SNR of the primary eclipses and the average SNR of the secondary eclipses to a value between zero and one, according to:
\begin{equation}
    A_0 = \frac{2}{\pi} arctan \left(\frac{\overline{\text{SNR}_{\text{p}}} + \overline{\text{SNR}_{\text{s}}}}{40}\right),
    \label{eq:attr_0}
\end{equation}
where the value of 40 in the denominator was found to be optimal by trial-and-error. Subscript p and s indicate primary and secondary eclipses.

$A_1$ is the ratio between the number of found primary and secondary eclipses $N_{\text{found}}$ and the total number of possible eclipses $N_{\text{possible}}$ in the time span of the light curve: 
\begin{equation}
    A_1 = \frac{N_{\rm found}}{N_{\rm possible}} = \frac{N_{\text{p}} + N_{\text{s}}}{N_{\rm possible, \text{p}} + N_{\rm possible, \text{s}}}\;.
    \label{eq:attr_1}
\end{equation}
The possible eclipses are determined by computing eclipse times from the period and the time of the first primary or secondary eclipse, only accepting those eclipse times that have coverage by data.

$A_2$ measures the average slope $S$ of the eclipses divided by the median of the absolute first time derivative $D_1$ of the light curve outside of the eclipses. The minimum slope is taken between the left and right side of each eclipse. This value is transformed analogously to $A_0$ to a domain between zero and one:
\begin{equation}
    A_2 = \frac{2}{\pi} arctan \left(\frac{\overline{min\left(S_{\rm left}, S_{\rm right}\right)}}{2.5\cdot median\left(\left| D_{1, \rm masked}\right|\right)}\right),
    \label{eq:attr_2}
\end{equation}
where the value of 2.5 in the denominator was again found to be optimal by trial-and-error. Slopes are calculated by dividing the depth of the eclipse at ingress or egress by the width of the ingress or egress, approximating the ingress and egress with straight lines.

$A_3$ is a measure for how symmetric the eclipses are, by looking at the depths $H$\footnote{H for height, since D is taken for the derivatives.} and slopes on the left and right side of the eclipses. The difference in depth measured at eclipse ingress and at egress is divided by their sum and the same is done for the slope at these locations. This is averaged over all eclipses and transformed to a value between zero and one:
\begin{equation}
    A_3 = \frac{1}{2} \left(\frac{1}{2} + \frac{1}{0.7} \overline{\left(\frac{2 \left|S_{\rm left} - S_{\rm right}\right|}{S_{\rm left} + S_{\rm right}}\right)} \cdot \overline{\left(\frac{2 \left|H_{\rm left} - H_{\rm right}\right|}{H_{\rm left} + H_{\rm right}}\right)}\right)^{-1},
    \label{eq:attr_3}
\end{equation}
if the second term between the outer brackets that starts with $1/0.7$, is below 0.5, it is multiplied by 
twice its complement with respect to 1. This has the effect of lowering the values of $A_3$ as it approaches one in a way that increases differentiation between values for light curves that would end up close to each other.

$A_4$ is a measure for how similar the eclipses are to each other, by taking the mean of the squared and mean-normalised differences from the mean for the SNR, depth ($H$), and width ($W$) values (for primary and secondary separately) and transforming them to a combined value between zero and one analogously to $A_3$:
\begin{equation}
    A_4 = \frac{1}{2} \left(\frac{1}{2} + 10 \cdot  B_{\text{SNR}} \cdot  B_{H} \cdot B_{W}\right)^{-1},
    \label{eq:attr_3}
\end{equation}
where each $B_i$ is calculated as follows: 
\begin{equation*}
    B_{W} = \frac{\sum_{\text{p}} \left( \frac{\left| W_{\text{p}} - \overline{W_{\text{p}}} \right|}{\overline{W_{\text{p}}}} \right)^2 + \sum_{\text{s}} \left( \frac{\left| W_{\text{s}} - \overline{W_{\text{s}}} \right|}{\overline{W_{\text{s}}}} \right)^2}{N_{\text{p}} + N_{\text{s}} - 1}\;.
\end{equation*}
Like $A_3$, if the second term, containing the $B_i$'s, is below 0.5, it is multiplied by 
twice its complement with respect to 1. 

\begin{figure}
\centering
\includegraphics[width=\hsize]{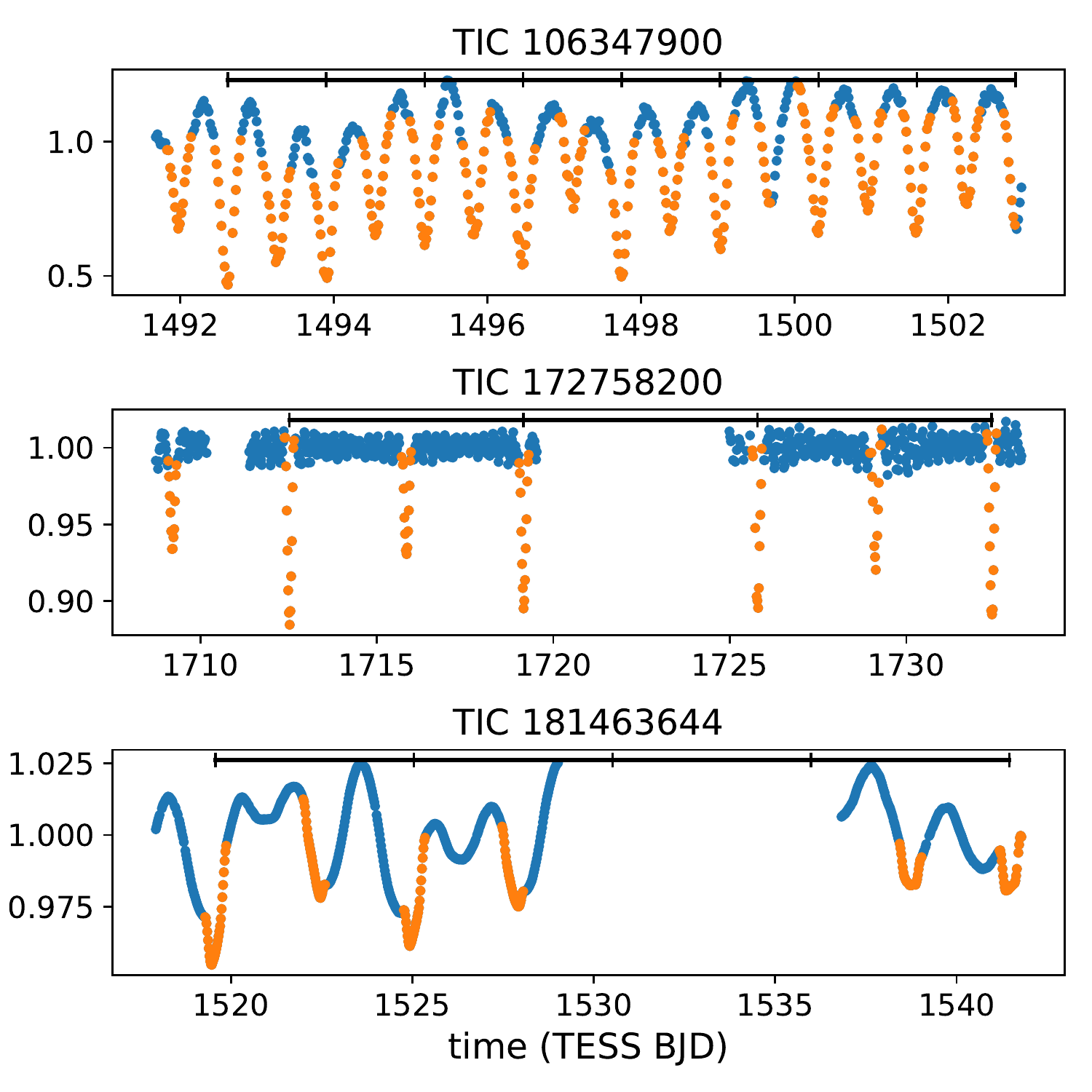}
  \caption{Part of the light curves of three examples taken to show the values of the $A_i$ formulae. Full light curves for these targets can be found in Figures\,\ref{fig:lcs}, \ref{fig:lcs-puls} and \ref{fig:lcs-var}.}
     \label{fig:lcs-values}
\end{figure}

Three light curves with different morphologies are chosen (see Fig.\,\ref{fig:lcs-values}) to give examples of the values obtained through these formulae. Only part of each light curve is shown here, since these examples can also be found in Appendix\,\ref{apx:examples}. Table\,\ref{tab:qtt-values} shows the values for each formula as well as the combined score.

\begin{table}
	\centering
	\caption{Values of $A_i$ for three examples.}
	\label{tab:qtt-values}
	\begin{tabular}{c l l l l l l}
	\hline
	TIC & $A_0$ & $A_1$ & $A_2$ & $A_3$ & $A_4$ & score \\
	\hline
	106347900 & 0.869 & 0.974 & 0.566 & 0.887 & 0.9999 & 0.430 \\
    172758200 & 0.927 & 0.938 & 0.831 & 0.945 & 0.9999 & 0.717 \\
    181463644 & 0.885 & 0.929 & 0.615 & 0.506 & 1.000 & 0.430 \\
	\hline
	\end{tabular}
	\tablefoot{Values of the $A_i$ formulae for the three examples shown in Fig.\,\ref{fig:lcs-values}, as well as their final scores.}
\end{table}

\section{Sky distribution plots}
\label{apx:sky-dist}

\begin{figure*}
\resizebox{\hsize}{!}
    {\includegraphics[width=\hsize,clip]{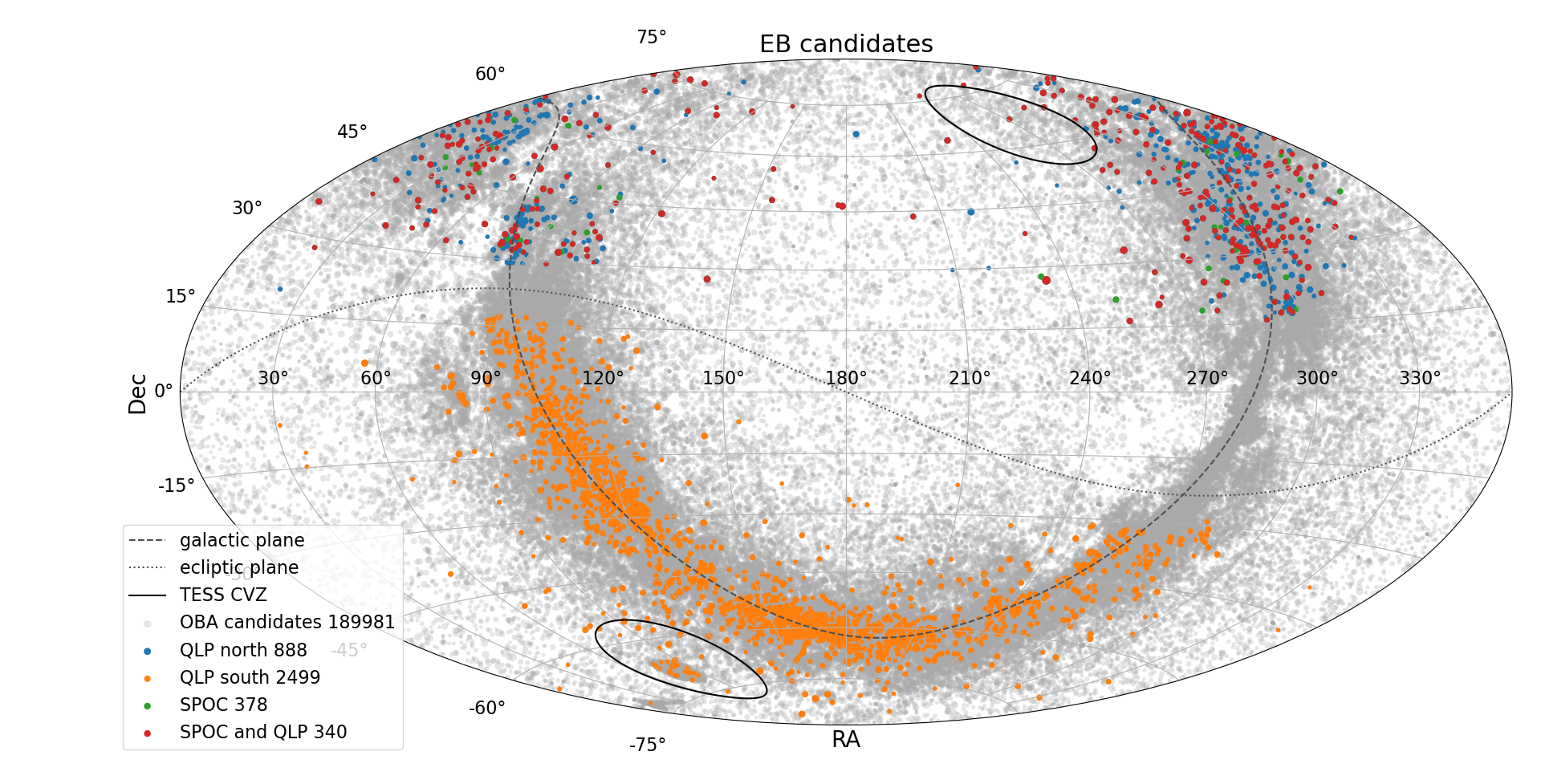}}
    \caption{Distribution on the sky of the overall sample (grey) and the EB candidates (other colours) in equatorial coordinates. The TESS CVZs are indicated with thick solid lines and the galactic plane is indicated by the dashed line. The dotted line traces the ecliptic plane.}
    \label{fig:equatorial}
\end{figure*}

\begin{figure*}
\resizebox{\hsize}{!}
    {\includegraphics[width=\hsize,clip]{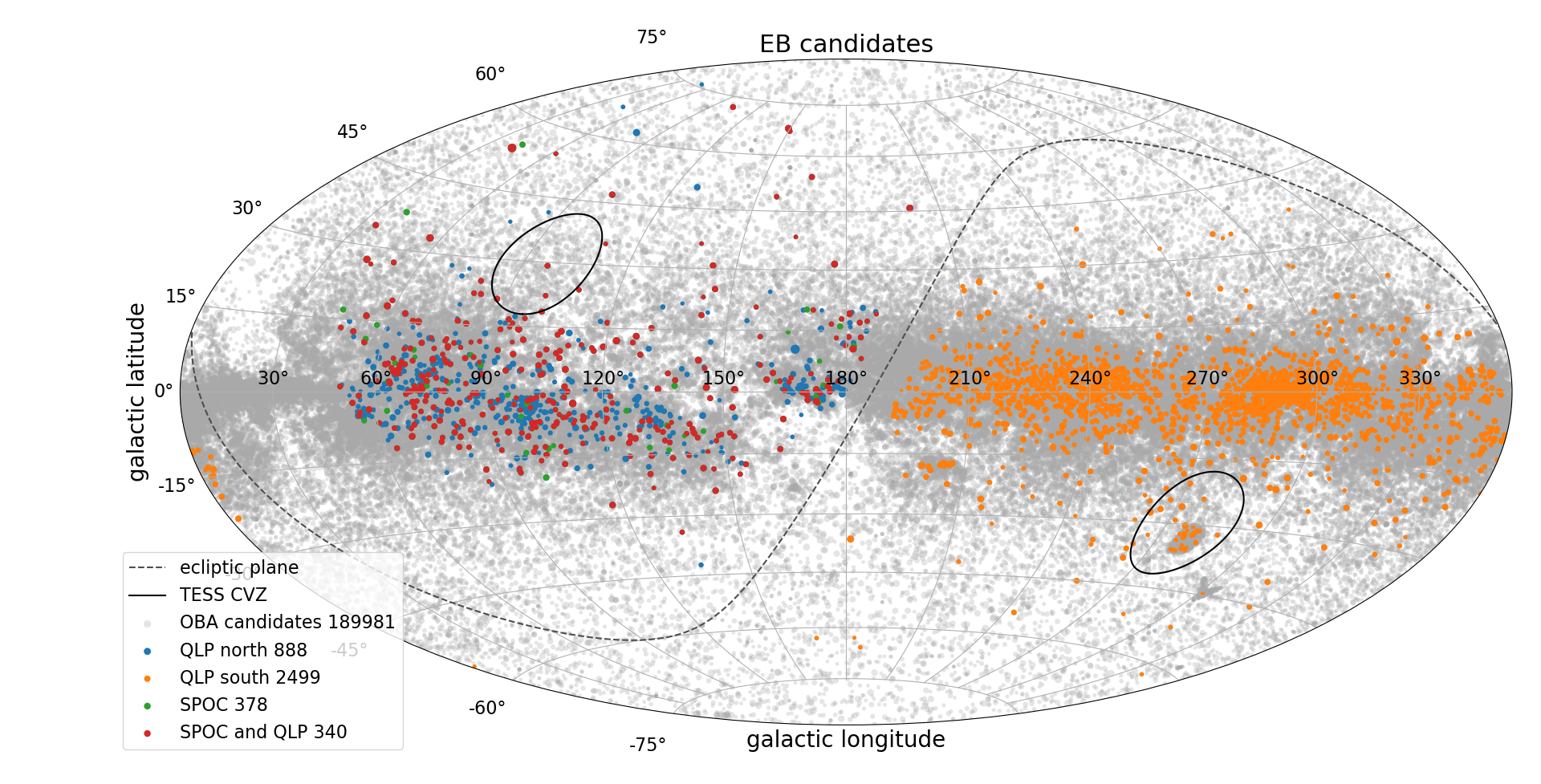}}
    \caption{Distribution on the sky of the overall sample (grey) and the EB candidates (other colours) in galactic coordinates. The TESS CVZs are indicated with thick solid lines and the ecliptic plane is indicated by the dashed line.}
    \label{fig:galactic}
\end{figure*}

\begin{figure*}
\resizebox{\hsize}{!}
    {\includegraphics[width=\hsize,clip]{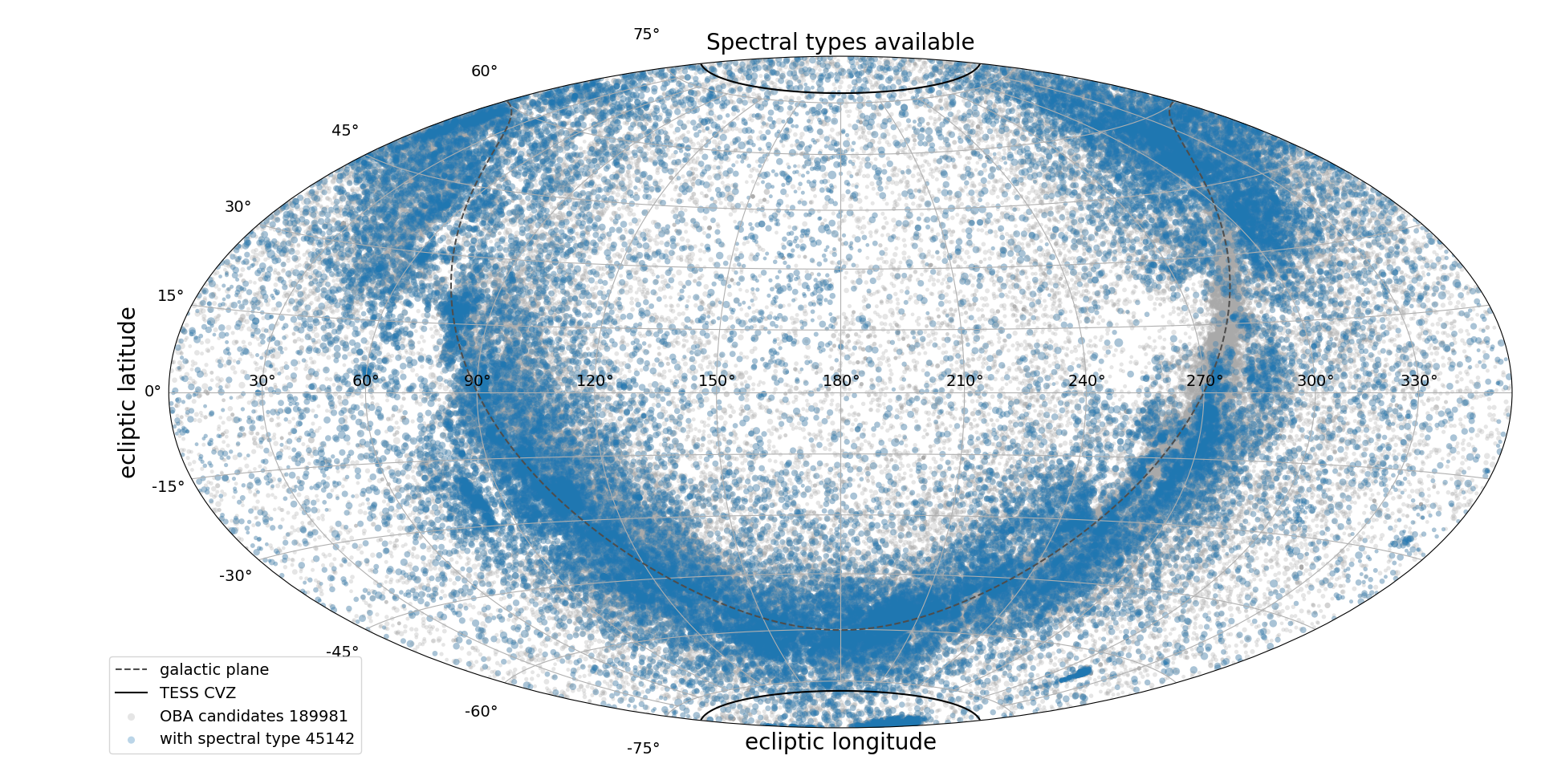}}
    \caption{Distribution of targets in the overall catalogue with \texttt{SIMBAD} spectral types available (blue). They cover the targets of the overall catalogue (grey) roughly in proportion to the target density.}
    \label{fig:simbad-sky}
\end{figure*}

\newpage

\section{Example light curves}
\label{apx:examples}

\begin{figure*}
\resizebox{\hsize}{!}
    {\includegraphics[width=\hsize,clip]{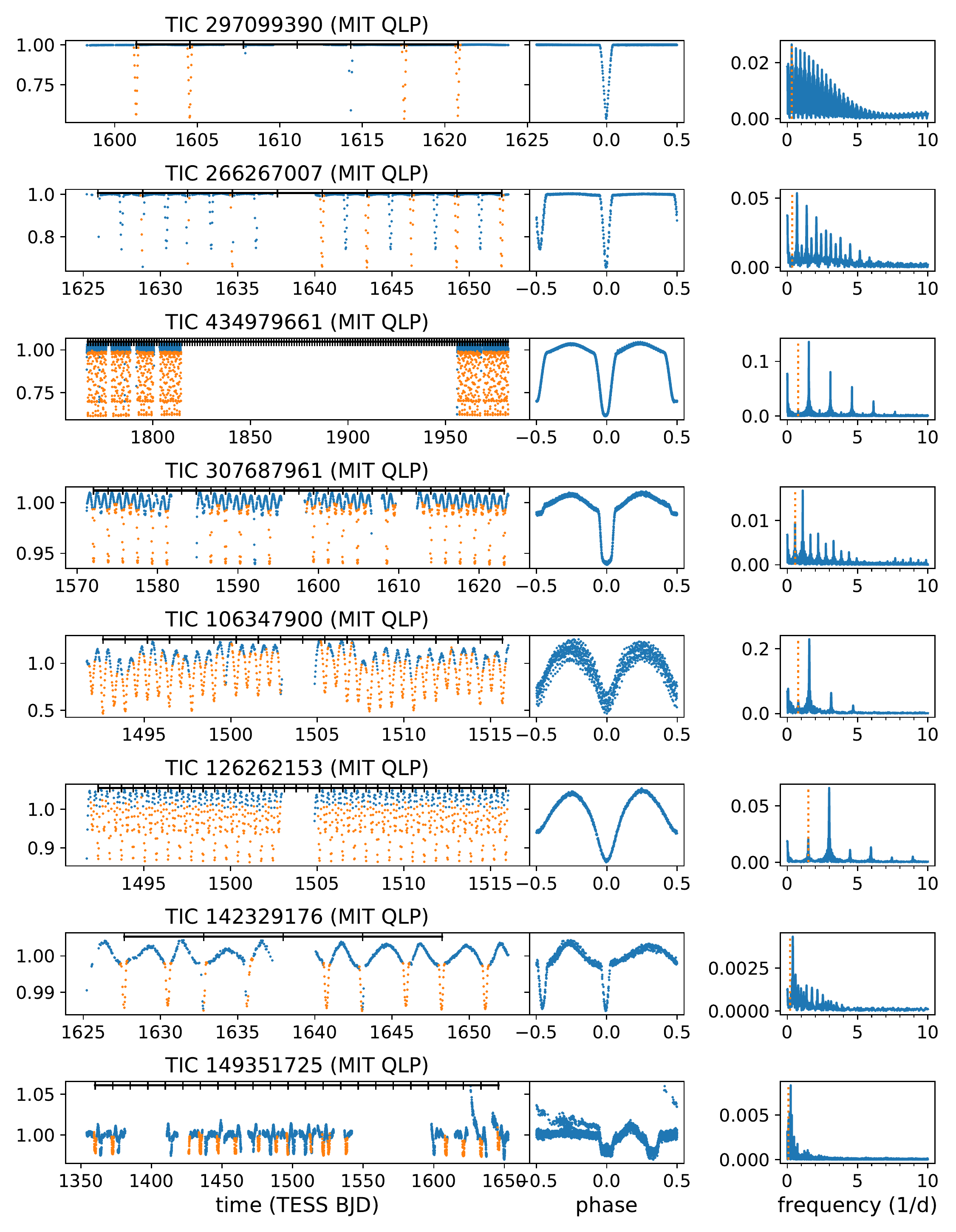}}
    \caption{Examples of EB candidates with no or limited additional variability, covering different light curve morphologies. The tick marks on the black line above the light curve indicates the positions of the (primary) eclipses. The vertical dotted line in the panels on the right indicates the orbital frequency.}
    \label{fig:lcs}
\end{figure*}

\begin{figure*}
\resizebox{\hsize}{!}
    {\includegraphics[width=\hsize,clip]{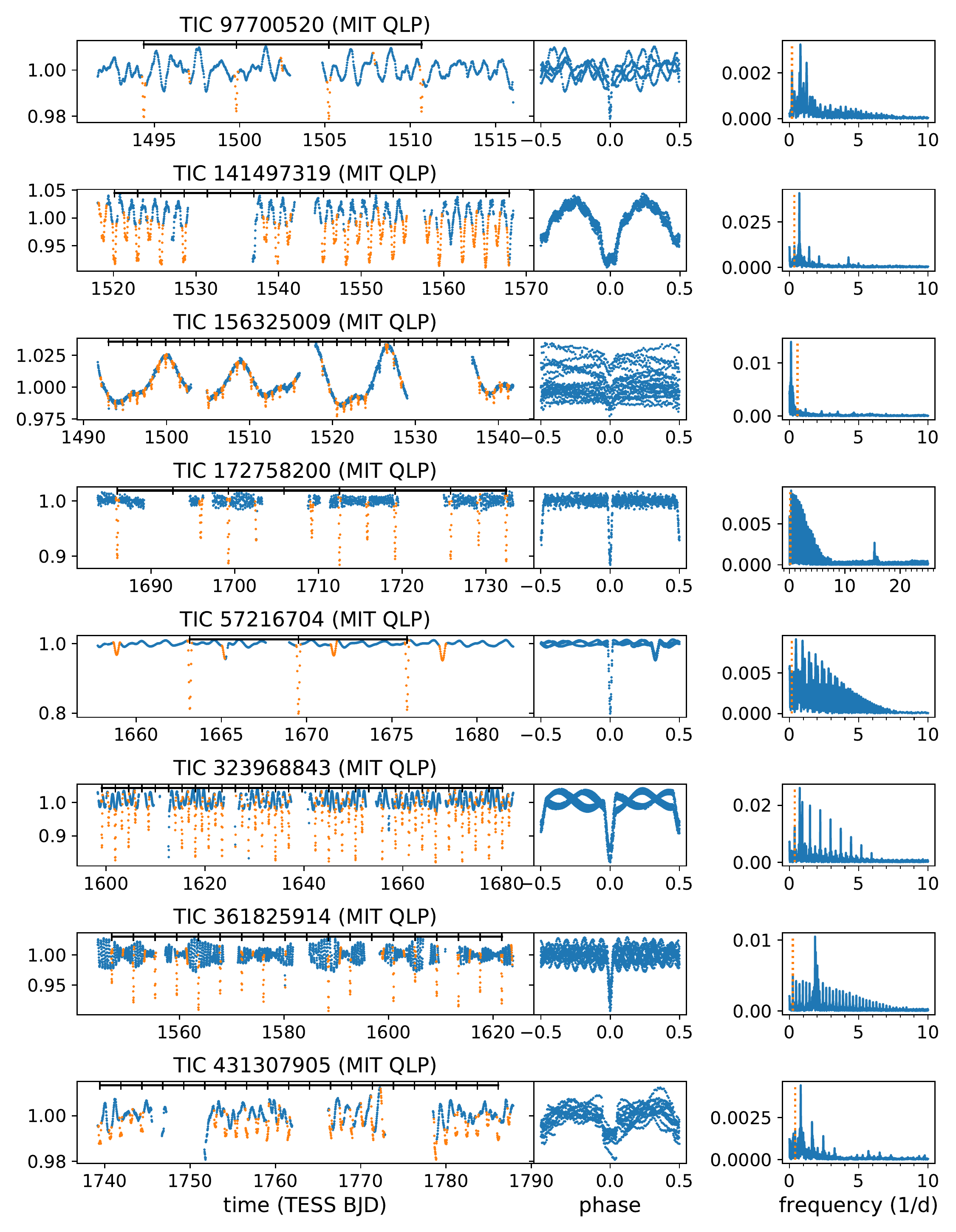}}
    \caption{Examples of EB candidates that have a clear presence of intrinsic variability, showing different combinations of most likely stellar pulsations and/or rotational modulation. The tick marks on the black line above the light curve indicates the positions of the (primary) eclipses. The vertical dotted line in the panels on the right indicates the orbital frequency.}
    \label{fig:lcs-puls}
\end{figure*}

\begin{figure*}
\resizebox{\hsize}{!}
    {\includegraphics[width=\hsize,clip]{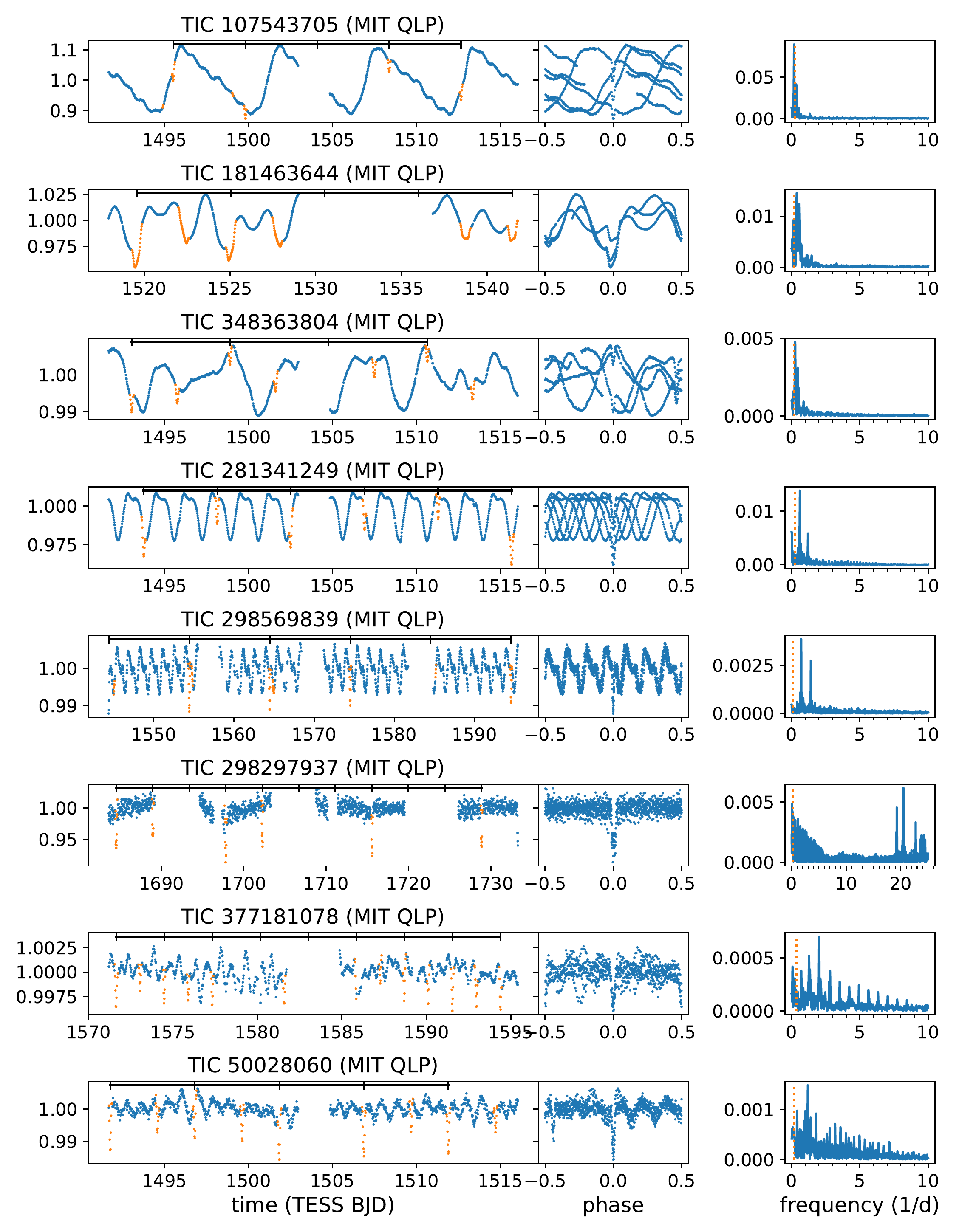}}
    \caption{Examples of EB candidates where the amplitude of any additional variability is dominant over the eclipse signal; typically difficult cases for automated detection. The tick marks on the black line above the light curve indicates the positions of the (primary) eclipses. The vertical dotted line in the panels on the right indicates the orbital frequency.}
    \label{fig:lcs-var}
\end{figure*}

\begin{figure*}
\resizebox{\hsize}{!}
    {\includegraphics[width=\hsize,clip]{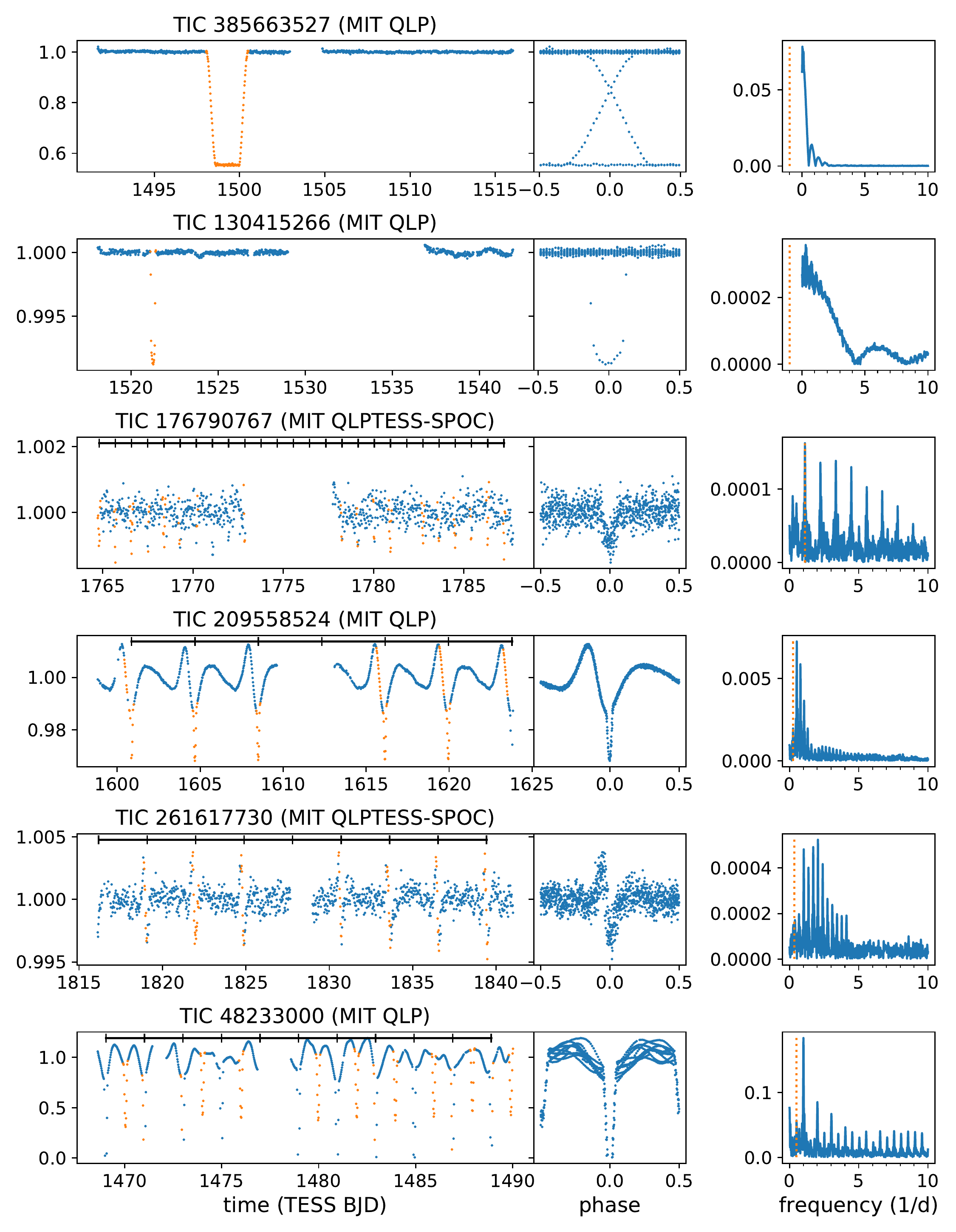}}
    \caption{Examples of some notable or interesting targets, including two light curves with a single eclipse and two heartbeat signals. The vertical dotted line in the panels on the right indicates the orbital frequency: a value of minus one means only a single eclipse is present. The bottom row represents a subset of cases where the data reduction resulted in primary eclipses that are cut off at zero, while they never reach their minima.}
    \label{fig:lcs-special}
\end{figure*}

Figures \ref{fig:lcs}, \ref{fig:lcs-puls}, \ref{fig:lcs-var} and \ref{fig:lcs-special} show various light curves of the OBA-type EB candidates selected by our algorithm as examples. The first set of targets has the eclipse signal as dominant or only signal and is chosen to cover some of the different possible light curve morphologies. The light curves are shown in the left panels with the eclipses marked in orange as they were marked by \texttt{ECLIPSR}. The line with vertical markers above the light curve indicates the positions of the primary eclipses. A phase-folded version of the light curve is shown to the right for a better view of the eclipse shapes and the right-most panels show a Lomb-Scargle periodogram. Figure\,\ref{fig:lcs-puls}, shows light curves with eclipses and a form of periodic variability intrinsic to the star, most likely associated with stellar pulsations. These are chosen to explore some of the different possibilities in combinations of eclipses and intrinsic signal most likely associated with pulsations: TIC 97700520 in the top row has an irregular-type signal that does not fold onto itself at the eclipse period, in TIC 323968843 the intrinsic signal is (close to) a multiple of the orbital period. TIC 141497319 has a recognisable boxy eclipse when folded and is modulated by intrinsic variability at a lower amplitude, while the folded light curve of TIC 431307905 in the bottom row looks distorted with the higher amplitude intrinsic signal relative to the eclipses.

Figure\,\ref{fig:lcs-var} presents some examples of eclipse candidates where the additional light variation is of similar or higher amplitude than the eclipses themselves, showing the advantage of searching in the time domain instead of in the frequency domain. Some of these show periodic variations while others show (semi-)irregular variability, with different reasons for being `difficult' cases for automatic detection. TIC 181463644 (second row) shows strongly deformed eclipses, barely recognised by a human eye: their respective location in troughs, on crests or on slopes determines the difference between the slope of the light curve before and after the eclipse versus that of the eclipse ingress and egress, thus increasing or decreasing the peak strength in the time derivatives. The bottom four rows show fast variation relative to the eclipse period where especially TIC 298569839 and TIC 377181078 (fourth- and second-to-last rows) also have shallow eclipses in both the absolute and relative sense. This combination in particular is tricky because it provides large `noise' peaks in the derivatives. 

The last set of light curves (Fig.\,\ref{fig:lcs-special}) is selected as notable or interesting examples. TIC 385663527 and TIC 130415266 show examples of single eclipses that got a high enough score to be included in the EB candidates (scores of 0.772 and 0.396, respectively). TIC 176790767 is a case that is difficult to draw conclusions about just from the light curve since the low eclipse depths leaves open the options of blending from a nearby source or an exoplanet. This is however a good example of the performance of \texttt{ECLIPSR} in low S/N, although any lower than this is likely problematic. TIC 209558524 and TIC 261617730 are examples of heartbeat stars where the latter reveals a rather ambiguous eclipse detection. Finally, TIC 48233000 is shown to illustrate a subset of data reduction artefacts that is present in the MIT QLP data set. The y-scale of the light curve is from one all the way to zero, and from the folded curve it is apparent that the primary eclipses are cut off at zero and do not reach their full depth. 

\end{appendix}

\end{document}